\documentclass[authoryear,final,5p,times,twocolumn]{elsarticle}

\usepackage[utf8]{inputenc}

\usepackage{amssymb}
\usepackage{amsmath}
\usepackage{multirow}

\usepackage{subcaption}
\usepackage{makecell}
\usepackage[authoryear]{natbib}
\usepackage{tabularx}

\usepackage[table,xcdraw]{xcolor} 
\usepackage{colortbl}             

\usepackage{tcolorbox}
\usepackage{hyperref}

\tcbuselibrary{most}
\tcbset{
    colback=white,
    colframe=black!60,
    boxrule=0.4pt,
    arc=1pt,
    left=3pt,
    right=3pt,
    top=2pt,
    bottom=2pt,
    fonttitle=\bfseries,
    before skip=3pt,
    after skip=3pt,
    enhanced,
    breakable=false,  
    width=\columnwidth}

\definecolor{gray1}{RGB}{150, 150, 150}  
\definecolor{gray2}{RGB}{190, 190, 190}
\definecolor{gray3}{RGB}{230, 230, 230}

\journal{Journal of Systems and Software}

\begin{document}

\begin{frontmatter}

\title{From Lab to Reality: A Practical Evaluation of Deep Learning Models and LLMs for Vulnerability Detection}
\tnotetext[accepted]{This manuscript has been accepted for publication in the Journal of Systems and Software.}

\author{Chaomeng Lu\corref{cor1}}
\ead{chaomeng.lu@kuleuven.be}

\author{Bert Lagaisse}
\ead{bert.lagaisse@kuleuven.be}

\cortext[cor1]{Corresponding author.}

\affiliation{
    organization={DistriNet Group-T, KU Leuven},
    city={Leuven},
    country={Belgium}
    }

\begin{abstract}
Vulnerability detection methods based on deep learning (DL) have shown strong performance on benchmark datasets, yet their real-world effectiveness remains underexplored. Recent work suggests that graph neural network-based and transformer-based models, including large language models (LLMs), yield promising results when evaluated on curated benchmark datasets. These datasets are typically characterized by similar data distributions and may contain synthetic samples, heuristic labels, or labeling noise. In this study, we systematically evaluate four representative DL models---Devign, ReVeal, LineVul, and VulBERTa---across four representative datasets: Juliet, Devign, BigVul, and ICVul. Each model is trained independently on each dataset, and the graph-based and CodeBERT representations adopted by these models are analyzed using t-SNE and centroid distance to examine vulnerability-related patterns. To assess realistic applicability, we further evaluate trained ReVeal and LineVul models, along with four open-weight LLMs, on VentiVul, our newly constructed temporally separated out-of-distribution (OOD) dataset comprising 200 recent vulnerabilities from Linux and Chromium.
Our experiments reveal that the evaluated representation methods struggle to distinguish vulnerable from non-vulnerable code and that trained models generalize poorly across datasets with differing distributions and characteristics. When evaluated on VentiVul, performance drops sharply, with most models failing to detect vulnerabilities reliably or distinguish vulnerable functions from their patched counterparts. These results expose a persistent gap between academic benchmarks and real-world deployment, emphasizing the value of our deployment-oriented evaluation framework and the need for more robust code representations, higher-quality datasets, and evaluation methods that account for vulnerability-fixing changes.

\end{abstract}

\begin{keyword}
Deep learning \sep Large Language Models \sep Vulnerability Detection \sep Code Representation
\end{keyword}

\end{frontmatter}

\section{Introduction}
Software vulnerabilities remain a persistent and critical threat in modern computing environments, enabling attackers to exploit bugs for unauthorized access, data leakage, or system compromise~\citep{binosi24the}. Over the years, numerous high-profile breaches have demonstrated that even a single vulnerability can lead to significant economic losses and reputational damage~\citep{zhou23sok}. As the scale and complexity of codebases increase, manually auditing source code for potential vulnerabilities becomes not only impractical but also error-prone~\citep{sun24gptscan}. In response, the research community and industry have turned to automation, leading to the development of a variety of techniques for detecting vulnerabilities automatically~\citep{neuhaus2009beauty,yan2017machine,zheng2020impact}. Rather than replacing static analysis, learning-based vulnerability detection can complement existing techniques by estimating the likelihood that previously unseen code contains vulnerabilities. These estimates can be used to prioritize potentially vulnerable functions or commits for further analysis or human review.

In recent years, deep learning models~\citep{shiri24a,chakraborty2021deep,fu2022linevul} and LLM-based approaches~\citep{zhang24prompt,lu2024grace,zhou25large} have achieved state-of-the-art results on widely used benchmark datasets such as Juliet~\citep{boland2012juliet}, Devign~\citep{zhou19devign}, and BigVul~\citep{fan2020ac}. These datasets have accelerated research progress by providing standardized evaluation environments that facilitate direct comparisons between different methods. However, these datasets suffer from issues such as limited size, low label quality, synthetic constructs, or duplication, which raise concerns about their representativeness~\citep{croft23dataquality,chen2023diversevul,croft23data}.

Our primary motivation arises from a mismatch between how learning-based vulnerability detectors are commonly evaluated and how they are expected to operate in practice. Although existing models report promising results on benchmark datasets, such results may not reliably indicate deployment effectiveness. In security applications, seemingly minor choices in dataset construction, experimental design, and evaluation can lead to unrealistic performance estimates and misleading conclusions about model capability~\citep{daniel22dos}. Moreover, an evaluation is practically meaningful only when its setting reflects the conditions under which the detector is intended to support developers~\citep{Francesco22just}. However, most vulnerability detection studies train and test models on samples drawn from the same historical dataset, where the data share similar distributions, labeling assumptions, class proportions, and project characteristics. Public datasets may further contain synthetic samples, heuristic labels, duplicated code, or unrelated functions labeled through commit-level assumptions~\citep{croft23dataquality,jimenez2019importance}. Consequently, strong within-dataset performance may reflect dataset-specific regularities rather than a transferable understanding of vulnerability-related code patterns~\citep{chakraborty2021deep,chakraborty24revisiting}. This motivates us to examine vulnerability detection from three complementary perspectives: whether the adopted code representations expose separable vulnerability-related signals, whether trained models generalize across heterogeneous datasets, and whether both DL models and LLMs can detect recent vulnerabilities and distinguish them from their patched counterparts under temporally separated, deployment-oriented evaluation.

To this end, we design a deployment-oriented evaluation framework for assessing vulnerability detection models, as shown in Figure~\ref{fig:framework}. We evaluate four widely studied DL-based models on four public vulnerability datasets. Our evaluation consists of three systematic stages. 
First, we examine whether the graph-based and token-based feature extraction pipelines adopted by representative detectors exhibit visible vulnerability-related structure before task-specific training. Second, we conduct within-dataset and cross-dataset evaluations to examine how DL model performance and transfer patterns vary across independently constructed benchmarks with different characteristics. Third, we evaluate trained ReVeal and LineVul models and four open-weight LLMs on VentiVul, a newly constructed, temporally separated out-of-distribution (OOD) benchmark containing recent Linux and Chromium vulnerabilities.

The novelty of this work lies in its systematic and deployment-oriented perspective. Our key contributions are summarized as follows:

\begin{itemize}
    \item We analyze graph-based and CodeBERT representations across four benchmark datasets, revealing limited and inconsistent separability between vulnerable and non-vulnerable code.
    \item We evaluate four DL models under within-dataset and cross-dataset settings, showing that dataset characteristics strongly affect model performance and transferability.
    \item We evaluate trained DL models and four open-weight LLMs on recent Linux and Chromium vulnerabilities, revealing substantial performance degradation and limited sensitivity to vulnerability-fixing changes.
    \item We construct VentiVul, a temporally separated OOD benchmark containing 200 CVEs and 263 vulnerable--patched function pairs, and introduce a deployment-oriented evaluation protocol that combines Whole-File and Function-Pair evaluation to distinguish vulnerability coverage from sensitivity to security patches.
\end{itemize}

Together, these contributions form a reproducible framework for evaluating the reliability, transferability, and robustness of vulnerability detection models under both controlled and deployment-oriented conditions. Overall, our findings provide strong empirical evidence that, despite recent progress, current DL-based vulnerability detection techniques remain insufficiently robust for real-world deployment. The evaluated LLMs also demonstrate limited and inconsistent vulnerability detection and patch-pair differentiation capabilities. All of our datasets, code, and results are available at the following link\footnote{\href{https://github.com/Chaomeng-Lu/A-Practical-Evaluation-of-Deep-Learning-Models-and-LLMs-for-Vulnerability-Detection.git}{https://github.com/Chaomeng-Lu/A-Practical-Evaluation-of-Deep-Learning-Models-and-LLMs-for-Vulnerability-Detection.git}}.

The rest of this paper is organized as follows. Section 2 reviews related work, Section 3 presents the evaluated datasets and VentiVul, Section 4 describes the methodology and experimental design, Section 5 details the results and analysis, Section 6 discusses our key findings and future research directions, Section 7 presents the threats and limitations, and Section 8 concludes the paper.

\section{Related Work}
This section summarizes prior work on DL and LLM-based vulnerability detection models and empirical studies evaluating their effectiveness.

\subsection{Vulnerability Detection Techniques}
Existing DL-based approaches have explored diverse code representations and model architectures~\citep{shiri24a}. Graph-based models such as Devign~\citep{zhou19devign}, DeepWukong~\citep{cheng21deepwu}, LineVD~\citep{hin22linevd}, DeepVD~\citep{wang23deepvd}, EPVD~\citep{zhang23vulnerability}, AMPLE~\citep{wen2023vulnerability}, and MGVD~\citep{qiu24vulnerability} incorporate structural semantics of source code using graph neural networks (GNNs) or convolutional neural networks (CNNs) to improve detection accuracy. Other methods like IVDetect~\citep{li21vulnera} and LIVABLE~\citep{wen24livable} emphasize interpretability and attempt to mitigate issues related to class imbalance. ReVeal~\citep{chakraborty2021deep} underscores a key challenge in the field—many of these models are trained on low-quality or biased datasets, which contributes to sharp performance degradation when evaluated in real-world scenarios. Although some studies, such as~\citep{hin22linevd} and~\citep{zhang23vulnerability}, have attempted real-world evaluations across different projects, their models were trained on imbalanced datasets and tested on historical codebases, limiting their ability to generalize to novel and emerging vulnerabilities.
Overall, the lack of robust cross-project and real-world validation remains a major barrier to the practical deployment of these techniques.
Another major direction uses pretrained language models to represent source code as token sequences. Models based on CodeBERT~\citep{feng2020codebert}, GraphCodeBERT~\citep{guo2020graphcodebert}, GPT-2~\citep{thapa22transformer}, and RoBERTa, such as VulBERTa~\citep{hanif22vulberta}, have been applied to vulnerability detection. LineVul~\citep{fu2022linevul} uses CodeBERT and attention information for function-level detection and line-level localization, while GRACE~\citep{lu2024grace} incorporates graph information into pretrained code representations. Transformer-based architectures have also been applied to vulnerability repair~\citep{fu24vision}.
More recently, LLMs such as GPT-3.5 and GPT-4 have shown potential for vulnerability detection, with existing studies examining prompt design and few-shot learning to improve their performance~\citep{zhou24large,zhang24prompt}. However, these approaches are still primarily evaluated on structured and curated datasets containing known vulnerabilities. Overall, the lack of robust cross-dataset and deployment-oriented evaluation remains a major barrier to determining whether current DL models and LLMs can generalize to recently discovered vulnerabilities.

\subsection{Empirical Studies}
Empirical studies have identified substantial limitations in DL-based vulnerability detection. Comparative analyses show that model architecture and preprocessing influence both effectiveness and efficiency~\citep{tang20a}. Different neural architectures also capture different code properties, with RNNs being effective at contextual information and CNNs at local patterns~\citep{lin2021deep}. Nevertheless, DL models do not consistently outperform simpler shallow-learning approaches~\citep{mazuera21shallow}, indicating that architectural complexity alone does not guarantee reliable vulnerability detection.
A central concern is the quality and distribution of vulnerability datasets. Unrealistic labeling assumptions can lead to sharp performance degradation under practical evaluation conditions~\citep{jimenez2019importance}. Similarly, state-of-the-art models may suffer performance drops exceeding 50\% due to flawed training data and simplistic model choices~\citep{chakraborty2021deep}. Since entirely noise-free vulnerability datasets are difficult to construct, noisy-label learning has been explored to improve model robustness to mislabeled samples~\citep{croft2022noisy}. Class imbalance also has a substantial effect, causing F1-score reductions of up to 73\% in some settings, although appropriate oversampling can mitigate part of this degradation~\citep{yang2023does}. These dataset-related issues further affect reproducibility and generalization. The reproduction of nine state-of-the-art models revealed considerable variation in their outputs and strong dependence on training-data characteristics~\citep{steenhoek23an}. Existing function-level models also struggle with multi-base-unit vulnerabilities, where the relevant evidence spans multiple code units~\citep{sejfia24toward}. More realistic evaluations confirm these limitations. Real-Vul exposes performance drops of up to 91 F1 points for models including DeepWukong, LineVul, ReVeal, and IVDetect~\citep{chakraborty24revisiting}. PRIMEVUL further reveals duplication, noisy labels, and unrealistic evaluation settings in existing datasets, showing that both DL models and advanced LLMs such as GPT-3.5 and GPT-4 can perform near randomly under real-world-like conditions~\citep{ding2024vulnerability}.

Distinct from earlier work, our study connects representation analysis, cross-dataset evaluation, and temporally separated OOD testing within a deployment-oriented framework. We examine how representation and dataset characteristics affect model behavior and evaluate both trained DL models and open-weight LLMs on recent vulnerabilities using Whole-File and Function-Pair evaluation.

\section{Datasets}

This section describes the benchmark datasets used in our experiments. We present four widely used vulnerability detection datasets and our newly curated, out-of-distribution, temporally separated evaluation dataset.

\subsection{Benchmark Datasets}

\textbf{Juliet}~\citep{boland2012juliet} is a large synthetic dataset developed by the National Institute of Standards and Technology (NIST). It contains small, labeled C/C++ code snippets with vulnerabilities corresponding to various CWE types. It is often used for benchmarking due to its controlled and well-labeled nature.

\textbf{Devign}~\citep{zhou19devign} contains C/C++ functions from open-source GitHub projects FFmpeg and QEMU, labeled as vulnerable or non-vulnerable manually. It offers a realistic but smaller dataset than BigVul. The dataset focuses on function-level vulnerability detection, making it suitable for fine-grained analysis and DL approaches.

\textbf{BigVul}~\citep{fan2020ac} is a large-scale dataset mined from commit histories of open-source C/C++ repositories. Vulnerable functions were identified based on rule-based matching of security-related commits, with buggy versions labeled as vulnerable, and both their fixes and unrelated unchanged functions labeled as non-vulnerable.

\textbf{ICVul}~\citep{lu2025icvul} is a high-quality software vulnerability dataset addressing challenges in label quality, diversity, and comprehensiveness. It focuses on CVEs linked to GitHub fix commits, extracting relevant functions, files, and metadata. It uses the SZZ algorithm to trace vulnerability-contributing commits and the ESC (Eliminate Suspicious Commit) technique to improve label reliability.

Table~\ref{tab:comparisondataset} presents a comparative overview of four selected datasets. Key characteristics include the number of functions (\textit{Functions}), the proportion of vulnerable samples (\textit{Vul Ratio}), availability of CWE type annotations (\textit{CWE Label}), and type of data collection method (\textit{Type}). The \textit{Type} column describes how vulnerability functions were collected: synthetically generated (Juliet), manually selected (Devign), or collected from real-world patch data (BigVul and ICVul).

\begin{table}[h]
\centering
\small
\caption{Comparative analysis of selected benchmark datasets.}
\label{tab:comparisondataset}
\begin{tabular}{lcccc}
    \hline
    \textbf{Dataset} & \textbf{Functions} & \textbf{Vul Ratio} & \textbf{CWE Label} & \textbf{Type}\\
    \hline
    Juliet & 253,002 & 37\% & YES & Synthetic \\
    Devign & 27,318 & 46\% & NO & Manual \\
    BigVul & 188,636 & 6\% & YES & Patches\\
    ICVul & 15,396 & 41\% & YES & Patches\\
    \hline
\end{tabular}
\end{table}

\subsection{VentiVul as an Out-of-Distribution Benchmark}

To further evaluate models that achieve promising results under controlled experimental settings, we constructed a deployment-oriented out-of-distribution benchmark named \textbf{VentiVul}. VentiVul contains recent vulnerabilities from two large and security-critical C/C++ projects, Linux and Chromium. We selected these projects because their publicly traceable vulnerability reports and fixing commits support reliable reconstruction of vulnerable–patched function pairs. To ensure data relevance and minimize labeling noise, the collection process was conducted manually, with human inspection applied to the CVE reports, linked issue reports, vulnerability-fixing commits, and affected source-code functions. For both Linux and Chromium, we collected CVEs disclosed starting from May 1, 2025, to ensure a clear temporal separation from the historical datasets used to train the four DL models evaluated in this study. The selected LLMs were also released before this cutoff. This temporal constraint reduces the likelihood that the evaluated vulnerabilities or their corresponding fixes were included in the training data of either the DL models or the LLMs.

We began with project-specific CVE records from the National Vulnerability Database (NVD), ordered them by disclosure date after the temporal cutoff, and manually inspected the records and their reference links in sequence. For each project, eligible CVEs were retained until the predefined target of 100 CVEs was reached. For Linux, these references typically led directly to the corresponding vulnerability-fixing commits. For Chromium, they commonly led to reports in the Chromium Issue Tracker, from which we traced the associated fixes and source-code changes. Despite this difference in the reference path, we applied the same inclusion and exclusion criteria to both projects. We retained only CVEs with valid and accessible references, explicit vulnerability-fixing changes in non-test C/C++ source files, and available CWE information. We excluded CVEs whose references were missing or inaccessible, whose fixes involved non-C/C++ code or only test files, whose changes contained excessive unrelated files or functions, or whose CWE information was unavailable. Each retained CVE and its associated code changes were then manually verified. The target size was fixed before data collection to obtain equally sized CVE subsets for Linux and Chromium; records were not selected based on model predictions or evaluation outcomes.

For each retained CVE, we extracted the affected functions before and after the corresponding fix. The before-fix version was labeled as vulnerable, while the after-fix version was labeled as non-vulnerable. The two versions were aligned using a pair identifier to form a vulnerable--patched function pair, with each pair containing exactly one vulnerable function and its corresponding patched function. When a CVE affected multiple functions, each affected function and its patched counterpart formed a separate pair. We additionally extracted functions from the same affected source files that were unrelated to the fix and labeled them as non-vulnerable. This construction preserves the substantial class imbalance encountered when vulnerability detectors are applied across all functions in affected source files, while the aligned pairs enable a focused evaluation of whether a model can distinguish vulnerable functions from their corresponding security patches. 
Table~\ref{tab:ventivul_statistics} summarizes the resulting dataset. VentiVul contains 13,326 functions associated with 200 CVEs and 263 strict one-to-one vulnerable--patched function pairs. The remaining extracted functions come from the same affected source files but are unrelated to the corresponding vulnerability-fixing changes. The Chromium subset contains 7,095 functions, 100 CVEs, 20 CWE categories, and 147 function pairs, while the Linux subset contains 6,231 functions, 100 CVEs, 27 CWE categories, and 116 function pairs. Overall, the benchmark covers 39 distinct CWE categories; this value is smaller than the sum of the project-level counts because some CWE categories occur in both projects.

\begin{table}[h]
\centering
\small
\caption{Statistics of the VentiVul benchmark.}
\label{tab:ventivul_statistics}
\begin{tabular}{l|rrrr}
\hline
\textbf{Project}
& \textbf{Functions}
& \textbf{CVEs}
& \textbf{CWEs}
& \textbf{Patch Pairs} \\
\hline
Chromium & 7,095  & 100 & 20 & 147 \\
Linux    & 6,231  & 100 & 27 & 116 \\
\hline
Overall  & 13,326 & 200 & 39 & 263 \\
\hline
\end{tabular}
\end{table}

\begin{figure}[h]
    \centering
    \begin{subfigure}[b]{\columnwidth}
        \centering
        \includegraphics[width=\textwidth]{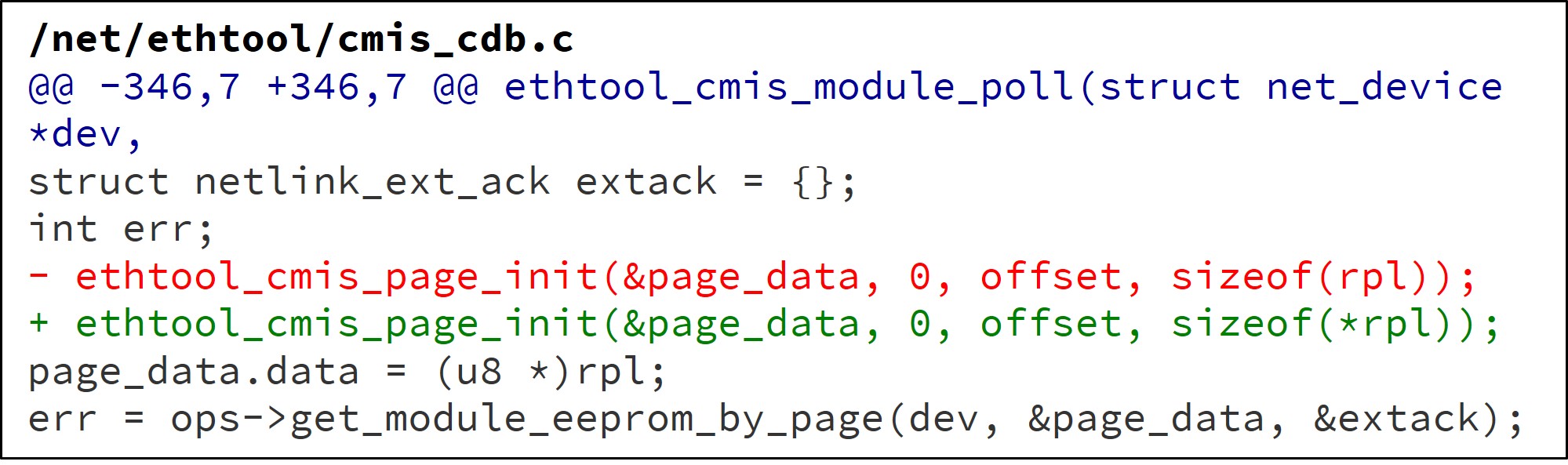}
        \caption{Fix for CVE-2025-37791: a single-function modification addressing stack corruption.}
        \label{fig:code_cve37791}
    \end{subfigure}
    \hfill
    \begin{subfigure}[b]{\columnwidth}
        \centering
        \includegraphics[width=\textwidth]{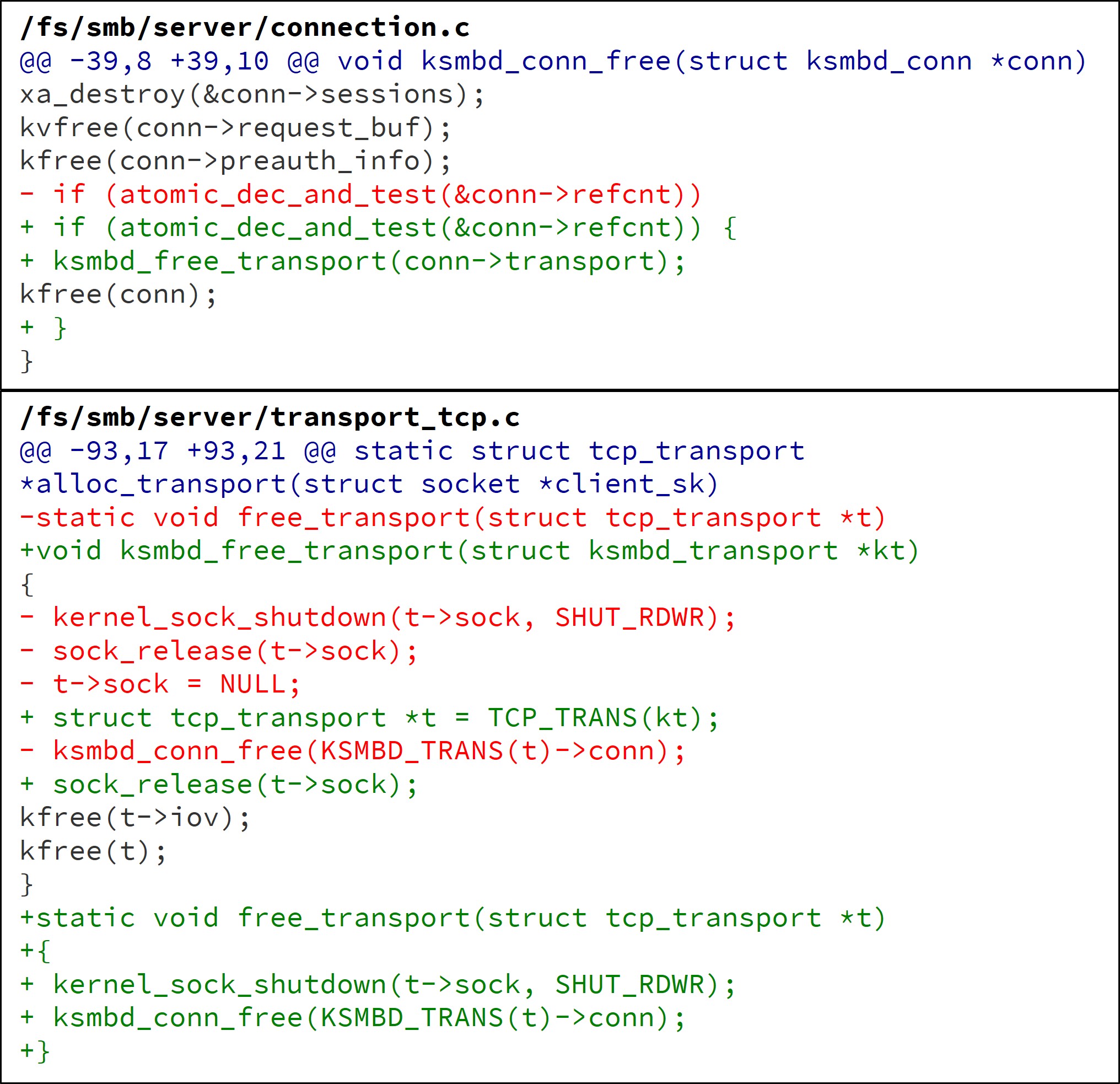}
        \caption{Fix for CVE-2025-37777: multi-function and multi-file modifications addressing a use-after-free vulnerability.}
        \label{fig:code_cve37777}
    \end{subfigure}
    \caption{Examples of vulnerability-fixing code modifications in the Linux subset of VentiVul.}
    \label{fig:code_cve}
\end{figure}

The collected CVEs cover different vulnerability types and code-change patterns. Figure~\ref{fig:code_cve} presents two examples from the Linux subset. Figure~\ref{fig:code_cve37791} shows CVE-2025-37791, for which the fix modifies a single function to address potential stack corruption. Figure~\ref{fig:code_cve37777} shows CVE-2025-37777, for which two functions across two files were modified to address a use-after-free vulnerability. These cases illustrate that a CVE may correspond to either a single function pair or multiple function pairs distributed across different files.

\section {Study Design}
\begin{figure*}[h]
    \centering
    \includegraphics[width=1\textwidth]{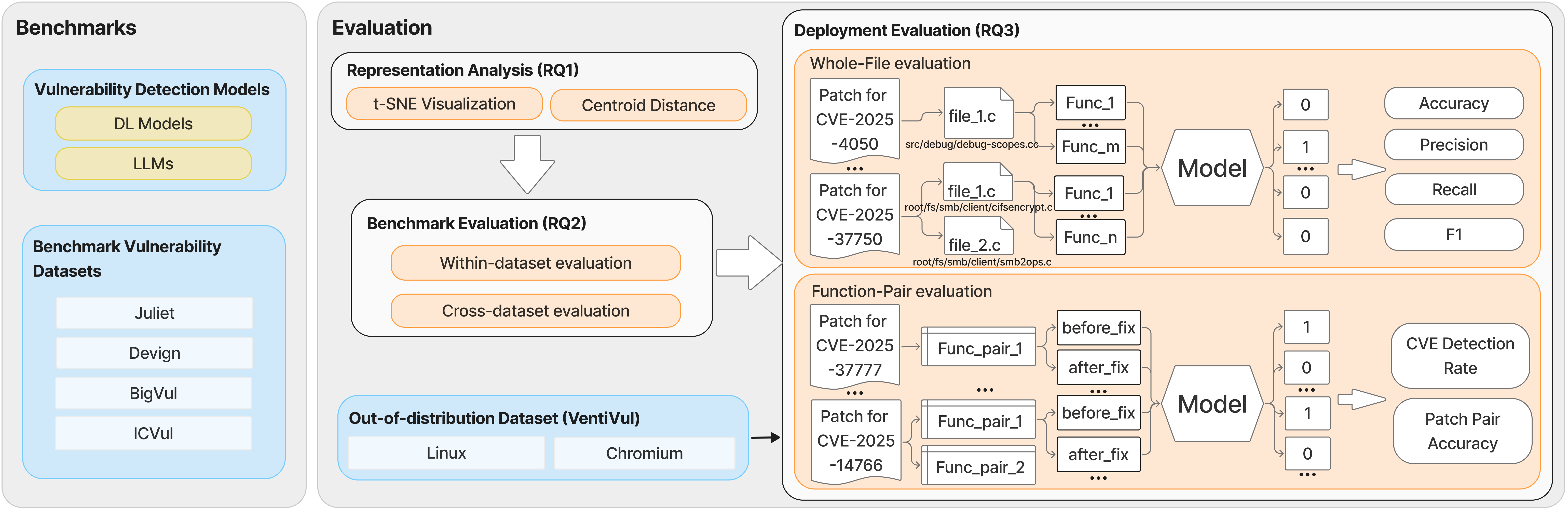}
    \caption{A deployment-oriented framework for evaluating vulnerability detection models.}
    \label{fig:framework}
\end{figure*}

To address our research questions, we propose a deployment-oriented evaluation framework for assessing vulnerability detection models, as illustrated in Figure~\ref{fig:framework}. The framework is organized from representation-level diagnostics to trained-model evaluation and finally to temporally separated OOD testing. This section first describes the benchmark model and dataset selection, then presents the research questions and experimental designs, and finally defines the evaluation metrics.

\subsection{Benchmark Selection}
To identify representative datasets and models, we surveyed vulnerability detection studies published in major software engineering, security, and machine learning venues, including ICSE, TSE, EMSE, TOSEM, JSS, MSR, FSE, PROMISE, RAID, IJCNN, and NeurIPS. We considered publications available up to May 2025, while acknowledging that the survey may not cover every relevant work.

\subsubsection{Model Selection}
The selection of DL models was guided by four criteria: (1) representativeness in prior vulnerability detection studies, (2) availability of code or sufficient implementation details, (3) compatibility with function-level C/C++ vulnerability detection, and (4) suitability for within-dataset, cross-dataset, and OOD evaluation. These criteria prioritize representative and reproducible models over exhaustive coverage of recently proposed architectures.
We categorized existing models based on the type of code representations they employ. While newer models exist, many continue to rely primarily on graph-based or token-based code representations. The models we considered include Devign~\citep{zhou19devign}, DeepWukong~\citep{cheng21deepwu}, ReVeal~\citep{chakraborty2021deep}, IVDetect~\citep{li21vulnera}, LineVul~\citep{fu2022linevul}, LineVD~\citep{hin22linevd}, VulBERTa~\citep{hanif22vulberta}, DeepVD~\citep{wang23deepvd}, AMPLE~\citep{wen2023vulnerability}, EPVD~\citep{zhang23vulnerability}, LIVABLE~\citep{wen24livable}, MGVD~\citep{qiu24vulnerability}, and GRACE~\citep{lu2024grace}. Broadly, these approaches can be grouped into two main categories:

a.~\textbf{Graph-based representations} are typically obtained using tools like \texttt{Joern}, which extract Abstract Syntax Tree (ASTs), Program Dependency Graphs (PDGs), and other graph-based structures. These representations capture the syntactic and control or data flow information of source code and are commonly used as input to GNNs. In many cases, token-level embeddings are generated by methods such as \texttt{Word2Vec}~\citep{mikolov2013distributed} or \texttt{GloVe}~\citep{pennington2014glove}. These embeddings are incorporated as node features within these graphs, enhancing the expressiveness of structural representations~\citep{wen2023vulnerability,li21vulnera}.

b. \textbf{Token-based representations}, which treat source code as a linear sequence of tokens and employ embedding techniques such as \texttt{Word2Vec} or pretrained transformer models like CodeBERT. These methods directly model the semantic and syntactic relationships between tokens and are widely used in LLMs for code understanding tasks~\citep{fu2022linevul,lu2024grace}.

From these categories, we selected four representative models for detailed evaluation:
\begin{itemize}
    \item \textbf{Devign}~\citep{zhou19devign}, which represents functions as code property graphs and applies a gated graph neural network to learn vulnerability-related structural features.
    \item \textbf{ReVeal}~\citep{chakraborty2021deep}, which combines \texttt{Joern}-derived program structures with token embeddings and addresses class imbalance in vulnerability datasets.
    \item \textbf{LineVul}~\citep{fu2022linevul}, which fine-tunes CodeBERT on function-level source code and uses self-attention scores for line-level vulnerability localization.
    \item \textbf{VulBERTa}~\citep{hanif22vulberta}, which pretrains a RoBERTa-based model on C/C++ source code using a code-specific tokenization pipeline.
\end{itemize}
Together, these models cover both graph-based and token-based vulnerability detection paradigms.

In addition, we included four open-weight LLMs (released before May 2025) to evaluate vulnerability detection without task-specific fine-tuning:
\begin{itemize}
    \item \textbf{Qwen2.5-Coder-32B-Instruct}~\citep{qwen25coder}, a 32B-parameter instruction-tuned code model released in November 2024 for code generation, reasoning, and repair.
    \item \textbf{DeepSeek-Coder-V2-Lite-Instruct}~\citep{deepseekcoderv2}, a code-oriented Mixture-of-Experts model released in June 2024 with 16B total and 2.4B activated parameters.
    \item \textbf{Qwen3-30B-A3B}~\citep{qwen3}, a Mixture-of-Experts model released in April 2025 with approximately 30B total and 3B activated parameters, supporting thinking and non-thinking inference modes.
    \item \textbf{Mistral-Small-24B-Instruct-2501}~\citep{mistralsmall3}, a 24B-parameter instruction-tuned model released in January 2025 for efficient instruction following, reasoning, and local inference.
\end{itemize}
These models cover both code-specialized and general-purpose architectures while supporting reproducible local deployment.

\subsubsection{Dataset Selection}
There are numerous publicly available datasets in the field of vulnerability detection. The datasets we considered include Juliet~\citep{boland2012juliet}, SARD~\citep{NISTSARD}, Devign~\citep{zhou19devign}, ReVeal~\citep{chakraborty2021deep}, BigVul~\citep{fan2020ac}, CrossVul~\citep{nikitopoulos21cross}, CVEFixes~\citep{bhandari21cve}, DiverseVul~\citep{chen2023diversevul}, D2A~\citep{zheng21d2a}, MegaVul~\citep{ni24megavul}, ICVul~\citep{lu2025icvul}. When selecting datasets for our pipeline, we consider two key factors. First, we prioritize datasets that are widely used and well-established in the research community \citep{croft23dataquality}. Second, we evaluate the structure and organization of the dataset to ensure it meets our methodological requirements. We aim to select datasets that offer diversity in terms of labeling strategy, data collection methodology, dataset volume, and class balance, ensuring a systematic evaluation of DL models. Based on these criteria, we selected Juliet, Devign, BigVul, and ICVul, which cover synthetic, manually labeled, and patch-derived data with substantially different dataset sizes and class distributions.

\subsection{Research Questions}
We organize the RQs from representation-level diagnostics to trained-model evaluation and finally to OOD testing.

\subsubsection{RQ1: Do current vulnerability-oriented code representation methods provide separable embeddings for vulnerable and non-vulnerable code before task-specific learning?}

\textbf{Motivation:}  
Learning-based vulnerability detectors rely on feature extraction pipelines that transform source code into graph-based or token-based representations before task-specific training. A downstream classifier can only learn from information preserved in these input representations. Therefore, if vulnerable and non-vulnerable functions are mapped to highly overlapping regions before training, the input representation may provide only limited directly separable signals to the downstream classifier. This does not mean that pre-training separability is sufficient for good detection performance; rather, it provides a diagnostic view of whether the adopted feature extraction method exposes potentially learnable vulnerability-related signals. This RQ deliberately does not evaluate the hidden representations learned after training ReVeal or LineVul. Instead, it isolates the vulnerability-label-free representations produced by the feature extraction pipelines adopted by these models. This distinction is important because our goal is to examine the representational substrate available to the downstream learner, rather than the end-to-end capability of a trained detector. We also analyze separability across vulnerability types when CWE metadata is available, because different CWE categories may correspond to different vulnerability mechanisms and code patterns.

\textbf{Experiment Design:}  
To investigate RQ1, we compare two feature extraction pipelines that correspond to the representation families used by the evaluated detectors: graph-based structural representations used by ReVeal-style models and token-based pretrained CodeBERT representations used by LineVul-style models. For each function, we extract a representation before vulnerability-specific training and analyze its separability with respect to vulnerability labels. For Devign and ICVul, we use all available samples for both visualization and quantitative analysis. For Juliet and BigVul, the full datasets are too large for readable visualization, so we use sampled subsets only for t-SNE plots. Specifically, we randomly sample 5\% of Juliet. For BigVul, whose original distribution is highly imbalanced, we construct a balanced visualization subset by including all vulnerable functions and randomly sampling the same number of non-vulnerable functions. This balancing is used only for visualization and does not affect model training or classification evaluation in RQ2.

We apply t-SNE to project high-dimensional representations into two dimensions for qualitative inspection. To complement visualization, we compute centroid distances between vulnerable and non-vulnerable samples in the normalized t-SNE space. When CWE metadata is available, we also compute centroid distances between selected vulnerability types and non-vulnerable samples.
These evaluations provide complementary descriptive evidence about whether the extracted features exhibit label- or CWE-related structure in the projected space. They are not used as direct measurements of separability in the original high-dimensional feature space or of downstream classification performance.

\subsubsection{RQ2: How do benchmark dataset characteristics influence within-dataset performance and cross-dataset generalization of DL-based vulnerability detectors?}

\textbf{Motivation:} 
Within-dataset evaluation is commonly used to report the performance of DL-based vulnerability detectors, but such evaluation may not reveal whether a model has learned vulnerability-relevant patterns or dataset-specific shortcuts. Cross-dataset evaluation provides additional insight into model transferability across independently constructed benchmarks. Vulnerability datasets differ substantially in data origin, project scope, class balance, labeling strategy, CWE coverage, and code similarity, and these differences jointly shape transfer patterns. To examine how model behavior varies under such heterogeneous benchmark conditions, we systematically evaluate models trained and tested across different datasets and interpret the observed transfer patterns in relation to the combined characteristics of the source and target datasets.

\textbf{Experiment design:} 
To investigate RQ2, we conduct within-dataset and cross-dataset evaluations involving both model- and dataset-level comparisons. Our goal is to examine how model performance changes when the source and target datasets differ in characteristics such as labeling strategy, class balance, dataset size, data origin, project scope, and code representation. Because these characteristics vary simultaneously across the selected datasets, the experiment is designed to reveal performance sensitivity under changing benchmark conditions.
We train four representative models, namely Devign, ReVeal, LineVul, and VulBERTa, independently on each of the four benchmark datasets using the model-specific settings recommended by their original implementations. 
The core of our experimental strategy is a cross-dataset evaluation paradigm. Each model is evaluated both on the dataset used for training and on the other independently constructed benchmark datasets. This setup allows us to observe how model performance changes when the target dataset differs from the training dataset in its distribution, construction procedure, labeling strategy, project composition, and class balance.
We also examine how models trained on different datasets perform on the same fixed target test set, enabling comparison of generalization patterns across varying training distributions.

To ensure data integrity and prevent leakage, we implemented several precautions. First, within each dataset, we performed strict train–test splits and filtered out duplicate functions, ensuring no overlap between training and test samples. Second, to guarantee evaluation consistency, for each target benchmark, we fixed its test set across all trained model configurations, ensuring that models evaluated on the same target dataset received identical test samples. Finally, to address temporal leakage, we introduced a time-wise out-of-distribution dataset, VentiVul, which was used in RQ3 for further evaluation.

\subsubsection{RQ3: How effectively do trained DL models and off-the-shelf LLMs detect recent vulnerabilities in a temporally separated deployment-oriented benchmark?}

\textbf{Motivation:} Conventional within-dataset evaluations draw the training, validation, and test sets from the same collection of historical vulnerabilities. However, commonly used vulnerability datasets may contain inaccurate labels and duplicated samples, which can affect model training and inflate evaluation results~\citep{croft23dataquality}. Moreover, strong performance under random or project-independent data splits does not necessarily demonstrate that a model can detect vulnerabilities disclosed after its training data were collected. Since vulnerability detectors are ultimately expected to identify previously unseen vulnerabilities, their effectiveness should also be evaluated on temporally separated samples from realistic software projects. This RQ therefore examines whether trained DL models and off-the-shelf LLMs can generalize to recent, security-critical vulnerabilities without additional task-specific training.

\textbf{Experiment Design:} To answer this RQ, we evaluated two trained DL model families, namely ReVeal and LineVul, in inference-only mode. These two models were selected as representative graph-based and pretrained transformer-based detectors, respectively, covering the two main DL representation families evaluated in RQ2.
These models were trained on the historical benchmark datasets used in RQ2 and were applied directly to VentiVul without retraining or adaptation. Because the ReVeal preprocessing pipeline cannot successfully process every function, its evaluation metrics are computed over the samples for which the model produces valid representations and predictions.
We additionally evaluated four open-weight LLMs: Qwen2.5-Coder-32B-Instruct, DeepSeek-Coder-V2-Lite-Instruct, Qwen3-30B-A3B, and Mistral-Small-24B-Instruct-2501. The LLMs were deployed locally and evaluated without parameter updates, thereby representing off-the-shelf use under reproducible inference conditions.
All models receive one function at a time. This function-level setting avoids variations caused by file length and context-window truncation and ensures that DL models and LLMs are evaluated on the same inputs. For each function, an LLM must return a binary vulnerability label. The response is constrained to a JSON object to enable deterministic parsing and evaluation.

Based on these function-level predictions, we consider two complementary evaluation settings, as summarized in Table~\ref{tab:rq3_settings}. The Whole-File setting evaluates all functions extracted from the affected source files, whereas the Function-Pair setting groups predictions by CVE and vulnerable--patched pair identifiers to evaluate CVE-level vulnerability coverage and patch-pair differentiation. CVE Detection Rate (CDR) treats a CVE as detected when at least one of its associated vulnerable functions is correctly classified. Patch Pair Accuracy (PACC) measures the proportion of valid vulnerable--patched function pairs for which the vulnerable function is classified as vulnerable and the corresponding patched function is classified as non-vulnerable.

\begin{table}[h]
\centering
\small
\caption{Evaluation Settings and metrics used for VentiVul.}
\label{tab:rq3_settings}
\begin{tabular}{p{2cm}|p{5.5cm}}
    \hline
    \textbf{Setting} & \textbf{Details} \\
    \hline
    Whole-File &
    All functions in the vulnerable files are evaluated. Each sample corresponds to a function. Functions modified by the vulnerability-fixing commit are labeled using their before-fix and patched status, while functions unrelated to the studied fix are treated as non-vulnerable with respect to that CVE. \\
    \hline
    Function-Pair &
    Predictions are grouped by CVE and vulnerable--patched pair identifiers to evaluate CVE-level vulnerability coverage and patch-pair differentiation using CDR and PACC. \\
    \hline
\end{tabular}
\end{table}

We considered three prompting settings for the LLMs, as summarized in Table~\ref{tab:rq3_prompt_settings}. These settings provide a controlled comparison of representative prompting strategies rather than an exhaustive or model-specific prompt search. In the \textit{zero-shot} setting, the model directly classifies the supplied function without demonstrations. In the \textit{few-shot} setting, the zero-shot query is preceded by a fixed set of labeled functions selected from complete vulnerable--patched pairs. In the \textit{chain-of-thought} setting, the prompt guides the model through security-relevant considerations before requesting the final decision. In all settings, the system prompt instructs the model to analyze only the supplied function, avoid assumptions about missing context, and return valid JSON only. Here, [X] denotes the target function. 

\begin{table*}[h]
\centering
\footnotesize
\caption{Prompting settings and prompts used for LLM evaluation on VentiVul.}
\label{tab:rq3_prompt_settings}
\begin{tabularx}{\textwidth}{l p{3.6cm} X}
    \hline
    \textbf{Setting} &
    \textbf{Description} &
    \textbf{Prompt} \\
    \hline

    Zero-shot &
    The LLM classifies a supplied function as vulnerable or non-vulnerable without demonstration examples. &
    \texttt{Determine whether the following function contains a security vulnerability. Return exactly one JSON object with this schema:
    \{"label":"vulnerable|non-vulnerable"\}. Function: [X]} \\
    \hline
    
    Few-shot &
    The LLM receives a fixed set of labeled functions from complete vulnerable--patched pairs before classifying the target function. &
    \texttt{Example $i$: [function]. Answer:
    \{"label":"vulnerable|non-vulnerable"\}. After all demonstrations, the model receives the same final query as in the zero-shot setting.} \\
    \hline
    
    Chain-of-thought &
    The LLM is instructed to consider attacker-controlled inputs, dangerous operations, missing validation, violated security conditions, and whether the supplied function provides sufficient evidence. &
    \texttt{Analyze the following function for security vulnerabilities. Consider: (1) attacker-controlled inputs; (2) dangerous memory, pointer, arithmetic, command, file, or resource operations; (3) missing validation or sanitization; (4) violated security conditions; and (5) whether the supplied function alone provides sufficient evidence. Make the best binary decision using only the supplied function. Return exactly one JSON object with the required label. Function: [X]} \\
    \hline
\end{tabularx}
\end{table*}

For few-shot evaluation, we select the same number of complete vulnerable--patched pairs from each project. A selected demonstration pair must contain exactly one vulnerable function and one patched function, ensuring that the demonstrations are balanced by both project and class. With $k=4$, for example, the prompt contains four demonstration functions obtained from two complete pairs, with one pair selected from each of two projects. The same fixed demonstrations are used across LLMs to support a controlled comparison.
To prevent leakage, all target samples sharing an isolation identifier with a selected demonstration are removed from the few-shot test set. For paired samples, the isolation identifier is constructed from the project and pair identifiers. Consequently, selecting one member of a vulnerable--patched pair excludes the entire pair from evaluation. We additionally verify that no target function has the same code hash as any demonstration. These controls prevent a function, its patched counterpart, or an identical code instance from appearing in both the few-shot prompt and the corresponding test set.

LLM outputs that cannot be parsed into a supported binary label are recorded as failed responses. We therefore record parsing success and apply two complementary evaluation policies. The \textit{valid-only} policy computes classification metrics using only successfully parsed predictions. For CDR, the valid-only policy includes only CVEs with at least one associated vulnerable function receiving a valid prediction. For PACC, it includes only structurally valid pairs for which all responses are successfully parsed. In contrast, the \textit{conservative} policy evaluates all samples and treats failed responses as incorrect predictions. Under this policy, a CVE is detected only when at least one of its vulnerable functions receives a valid positive prediction, while a patch pair is differentiated only when every function in the pair receives a valid and correct prediction. Reporting both policies separates classification performance on valid outputs from the practical impact of malformed or missing LLM responses.

\subsection{Implementation Details}
All LLM experiments were conducted locally using vLLM on a multi-GPU server equipped with NVIDIA B200 GPUs. We used deterministic decoding with temperature set to $0$, top-$p$ to $1$, a fixed random seed of $42$, and a maximum output length of 384 tokens. Qwen3-30B-A3B was evaluated with thinking mode disabled to maintain comparable inference and output settings across models. Each function was evaluated independently, and invalid or unparseable responses were retained as failed predictions rather than manually corrected or regenerated. All DL experiments were conducted on a multi-GPU server equipped with NVIDIA H100 GPUs. ReVeal, Devign, LineVul, and VulBERTa were implemented using the open-source code and recommended settings provided by their original authors. The trained ReVeal and LineVul checkpoints were applied directly to VentiVul without retraining or adaptation. Under both Whole-File and Function-Pair evaluation, DL models and LLMs received one function at a time to ensure a consistent input scope.

\subsection{Evaluation Metrics}

To comprehensively evaluate the performance of our models, we adopt a set of feature representation measures as well as classification and deployment metrics. These metrics allow us to assess both model predictions and the separability of code representations extracted before downstream classifier training.

\subsubsection{Feature Representation Measures}
To assess the separability of the graph-based and CodeBERT-based features extracted before downstream classifier training, we employ both visualization and quantitative analysis. We use t-Distributed Stochastic Neighbor Embedding (t-SNE) to reduce high-dimensional feature vectors into two dimensions for visualization. This technique enables us to inspect the separability of vulnerable and non-vulnerable code samples in the embedded space. Additionally, we calculate the centroid distance between classes based on the normalized t-SNE coordinates. Let $\mu_{vul}$ and $\mu_{non}$ denote the mean feature vectors (centroids) of the vulnerable and non-vulnerable samples, respectively. The centroid distance is defined as the Euclidean distance between the two centroids:
\begin{equation}
D_{\text{centroid}} = \left\| \mu_{\text{vul}} - \mu_{\text{non}} \right\|_2
\label{eq:centroid}
\end{equation}
This distance is computed in the normalized 2D space after applying Min-Max scaling to the t-SNE output. A larger centroid distance indicates greater geometric separation between class centroids in the normalized t-SNE projection. Because t-SNE does not preserve all global distances from the original high-dimensional space, this metric is interpreted as a complementary descriptive indicator rather than a direct measure of classification effectiveness.

\subsubsection{Classification and Deployment Metrics}

We report four standard metrics commonly used for binary classification tasks: accuracy, precision, recall, and F1-score. Let $TP$, $TN$, $FP$, and $FN$ denote the number of true positives, true negatives, false positives, and false negatives, respectively.

\textbf{Accuracy} measures the overall proportion of correct predictions and is defined as:
\begin{equation}
    \text{Accuracy} = \frac{TP + TN}{TP + TN + FP + FN}
    \label{eq:accuracy}
\end{equation}

\textbf{Precision} evaluates the proportion of correctly predicted positive instances among all instances predicted as positive:
\begin{equation}
    \text{Precision} = \frac{TP}{TP + FP}
    \label{eq:precision}
\end{equation}

\textbf{Recall}, also known as sensitivity, quantifies the proportion of actual positive instances that were correctly predicted:
\begin{equation}
    \text{Recall} = \frac{TP}{TP + FN}
    \label{eq:recall}
\end{equation}

\textbf{F1-score} is the harmonic mean of precision and recall and provides a balanced measure of the model's performance:
\begin{equation}
    \text{F1} = 2 \cdot \frac{\text{Precision} \cdot \text{Recall}}{\text{Precision} + \text{Recall}}
    \label{eq:f1}
\end{equation}

The four standard classification metrics are used under the Whole-File setting. Under the Function-Pair setting, we additionally report CDR for CVE-level vulnerability coverage and PACC for vulnerable--patched pair differentiation. \textbf{CVE Detection Rate (CDR)} measures the proportion of CVEs for which at least one associated vulnerable function is correctly identified:
\begin{equation}
    \text{CDR} =
    \frac{
        \sum_{c \in \mathcal{C}}
        \mathbb{I}
        \left[
            \exists f \in \mathcal{F}_{c} :
            \hat{y}_{f} = 1
        \right]
    }{
        |\mathcal{C}|
    }
    \label{eq:cacc}
\end{equation}
where $\mathcal{C}$ denotes the set of CVEs, $\mathcal{F}_{c}$ denotes the set of vulnerable functions associated with CVE $c$, and $\hat{y}_{f}$ is the predicted label of function $f$.

\textbf{Patch Pair Accuracy (PACC)} measures the proportion of valid vulnerable--patched function pairs for which the vulnerable function is correctly classified as vulnerable and the corresponding patched function is correctly classified as non-vulnerable:
\begin{equation}
    \text{PACC} =
    \frac{
        \sum_{p \in \mathcal{P}}
        \mathbb{I}
        \left[
            \hat{y}_{p}^{\mathrm{vul}} = 1
            \land
            \hat{y}_{p}^{\mathrm{patch}} = 0
        \right]
    }{
        |\mathcal{P}|
    }
    \label{eq:pacc}
\end{equation}
where $\mathcal{P}$ denotes the set of valid vulnerable--patched function pairs, and $\hat{y}_{p}^{\mathrm{vul}}$ and $\hat{y}_{p}^{\mathrm{patch}}$ denote the predicted labels of the vulnerable and patched functions in pair $p$, respectively.

\section{Results and Analysis}

In this section, we present the experimental results and summarize the key findings of our three RQs.

\subsection{Representation Analysis (RQ1)}

\subsubsection{Visual Separability}

\begin{figure*}[h]
\centering
\begin{subfigure}[b]{1\linewidth}
    \centering
    \includegraphics[width=1\linewidth]{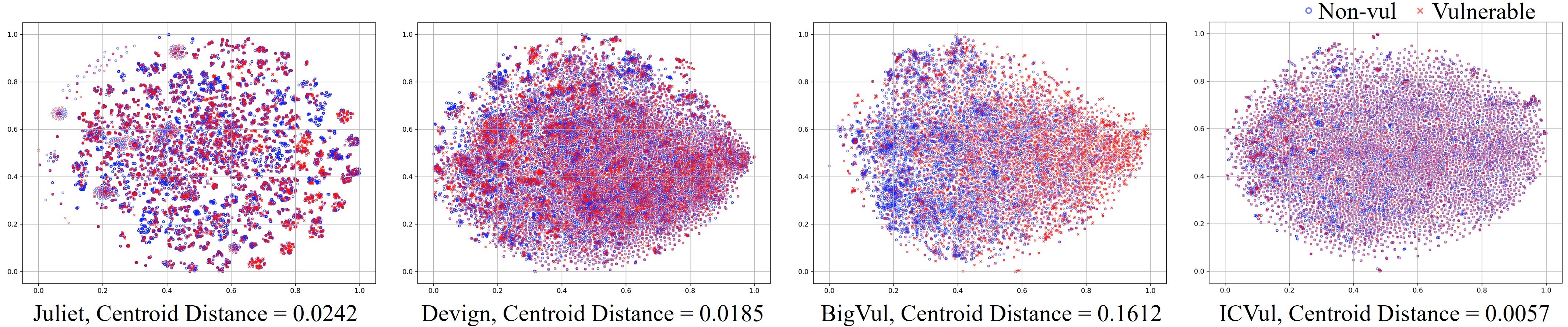}
    \caption{GNN: Feature space visualization and cluster analysis using t-SNE and centroid distances.}
    \label{fig:gnn_tsne}
\end{subfigure}
\begin{subfigure}[b]{1\linewidth}
    \centering
    \includegraphics[width=1\linewidth]{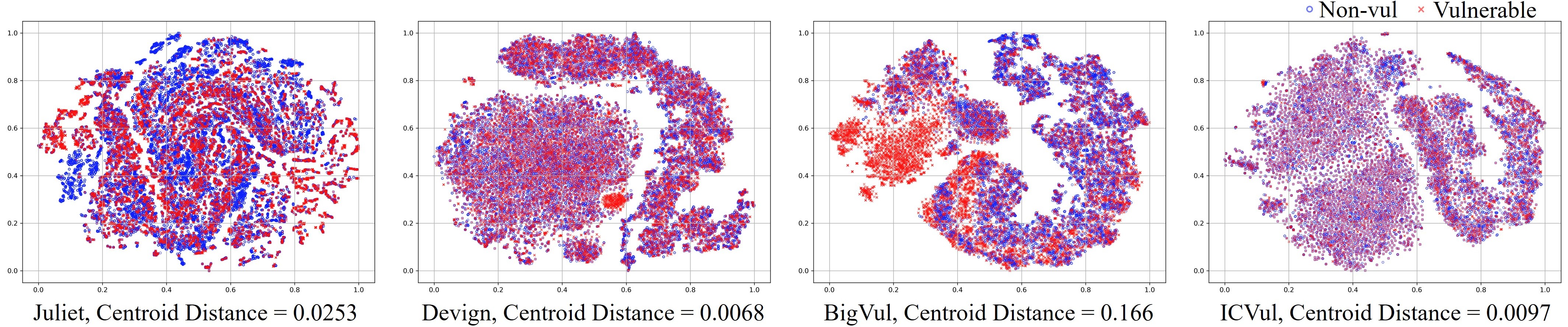}
    \caption{CodeBERT representations: Feature space visualization and cluster analysis using t-SNE and centroid distances.}
    \label{fig:codebert_tsne}
\end{subfigure}
\caption{Comparison of feature space visualizations and cluster analysis on four benchmark datasets using (a) GNN and (b) CodeBERT representations.}
\label{fig:tsne_comparison}
\end{figure*}

Figure~\ref{fig:gnn_tsne} and~\ref{fig:codebert_tsne} show binary t-SNE plots of features extracted before downstream classifier training on four benchmark datasets, where each point represents a function colored by its label (Non-vulnerable or Vulnerable).
The feature space visualizations using GNN-based embeddings (Figure \ref{fig:gnn_tsne}) show limited class separability across most benchmark datasets. In particular, Juliet and ICVul exhibit heavy overlap between vulnerable and non-vulnerable functions, indicated by their small centroid distances (0.0242 and 0.0057, respectively). Although BigVul displays a larger centroid distance (0.1612), the clusters are still not clearly distinct. This suggests that even though graph-based representations such as AST and PDG are rich in structural information, they do not necessarily translate into discriminative features for vulnerability classification when visualized in low-dimensional space. The CodeBERT-based embeddings (Figure~\ref{fig:codebert_tsne}) demonstrate better, but still imperfect, class separation on some datasets. BigVul shows relatively more pronounced clustering of vulnerable and non-vulnerable samples, with a centroid distance of 0.166, while Devign still has a small centroid distance of 0.0068. However, the separation is still inconsistent across datasets. 
ICVul and Juliet, whose construction procedures are intended to provide more controlled or carefully filtered labels, do not exhibit strong separation. This suggests that transformer-based methods can sometimes better capture the contextual cues of vulnerabilities, but their effectiveness is still limited in realistic, well-labeled datasets. From Figure \ref{fig:tsne_comparison}, we can find that across all datasets, both methods struggle to form well-separated clusters of vulnerable and non-vulnerable functions. This effect is especially prominent in ICVul and Devign, which contain real-world functions with substantial syntactic and semantic heterogeneity.

\subsubsection{Quantitative Separability}

\begin{figure}[h]
\centering
\begin{subfigure}[b]{1\linewidth}
    \centering
    \includegraphics[width=\linewidth]{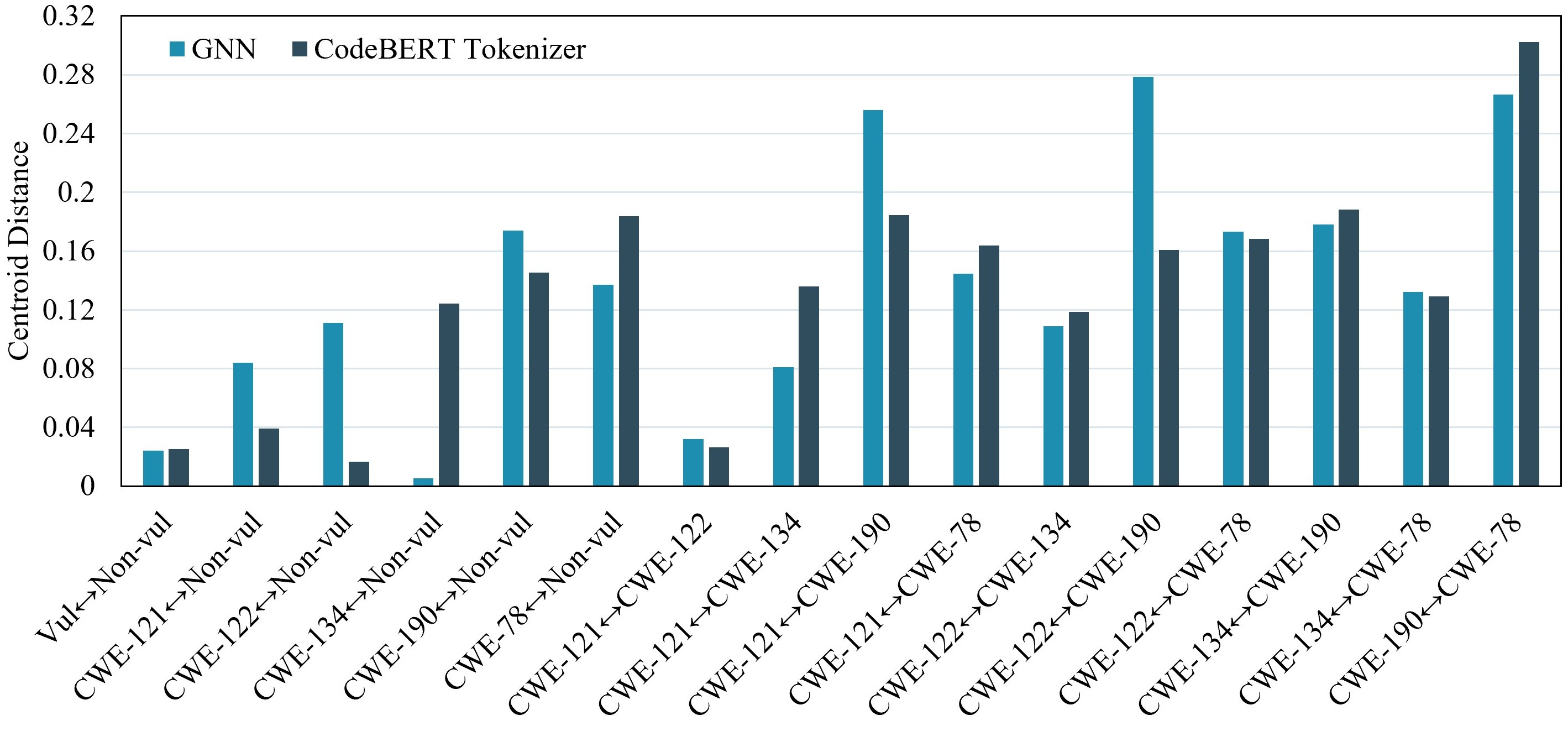}
    \caption{Juliet}
    \label{fig:juliet_cd}
\end{subfigure}
\hfill
\begin{subfigure}[b]{1\linewidth}
    \centering
    \includegraphics[width=\linewidth]{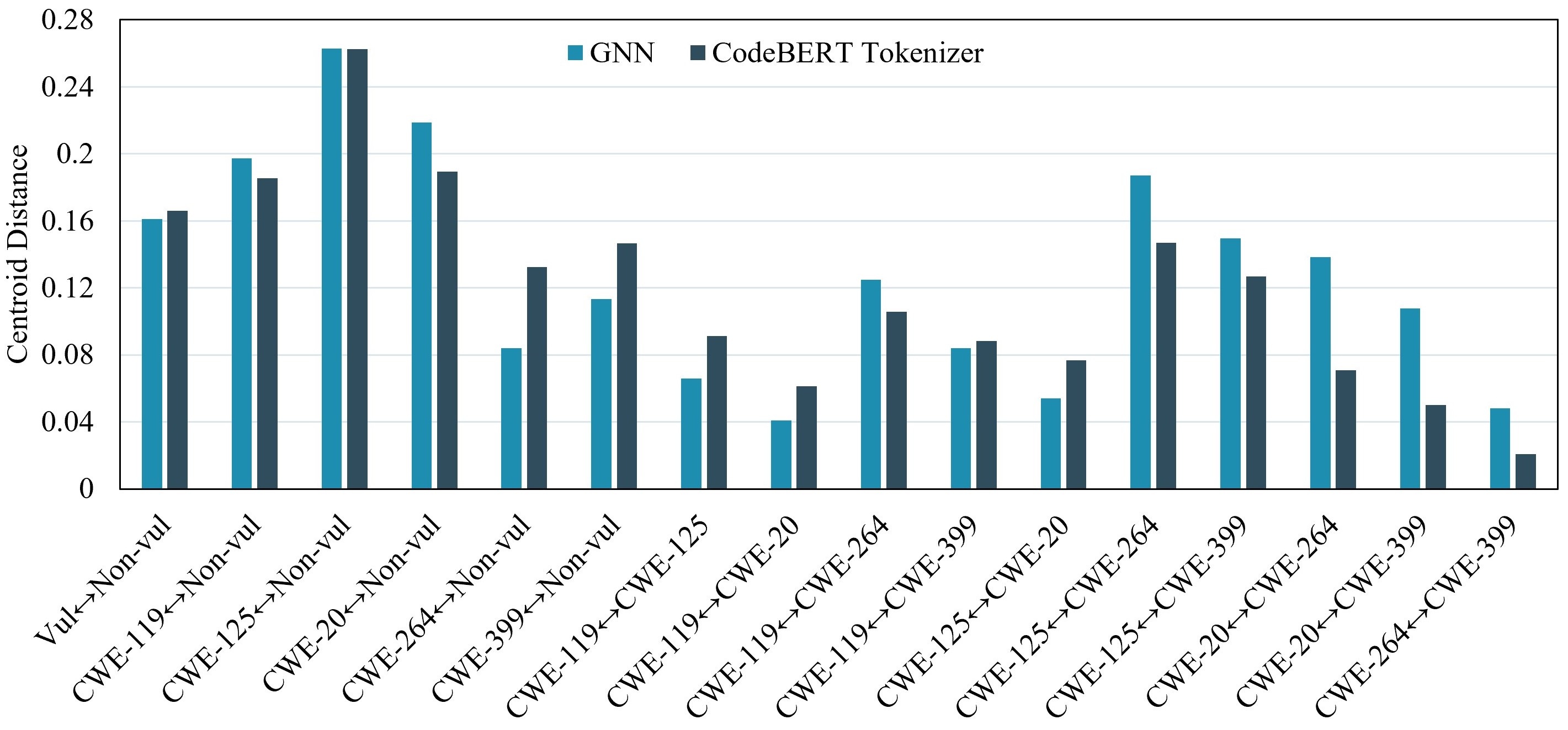}
    \caption{BigVul}
    \label{fig:bigvul_cd}
\end{subfigure}
\hfill
\begin{subfigure}[b]{1\linewidth}
    \centering
    \includegraphics[width=\linewidth]{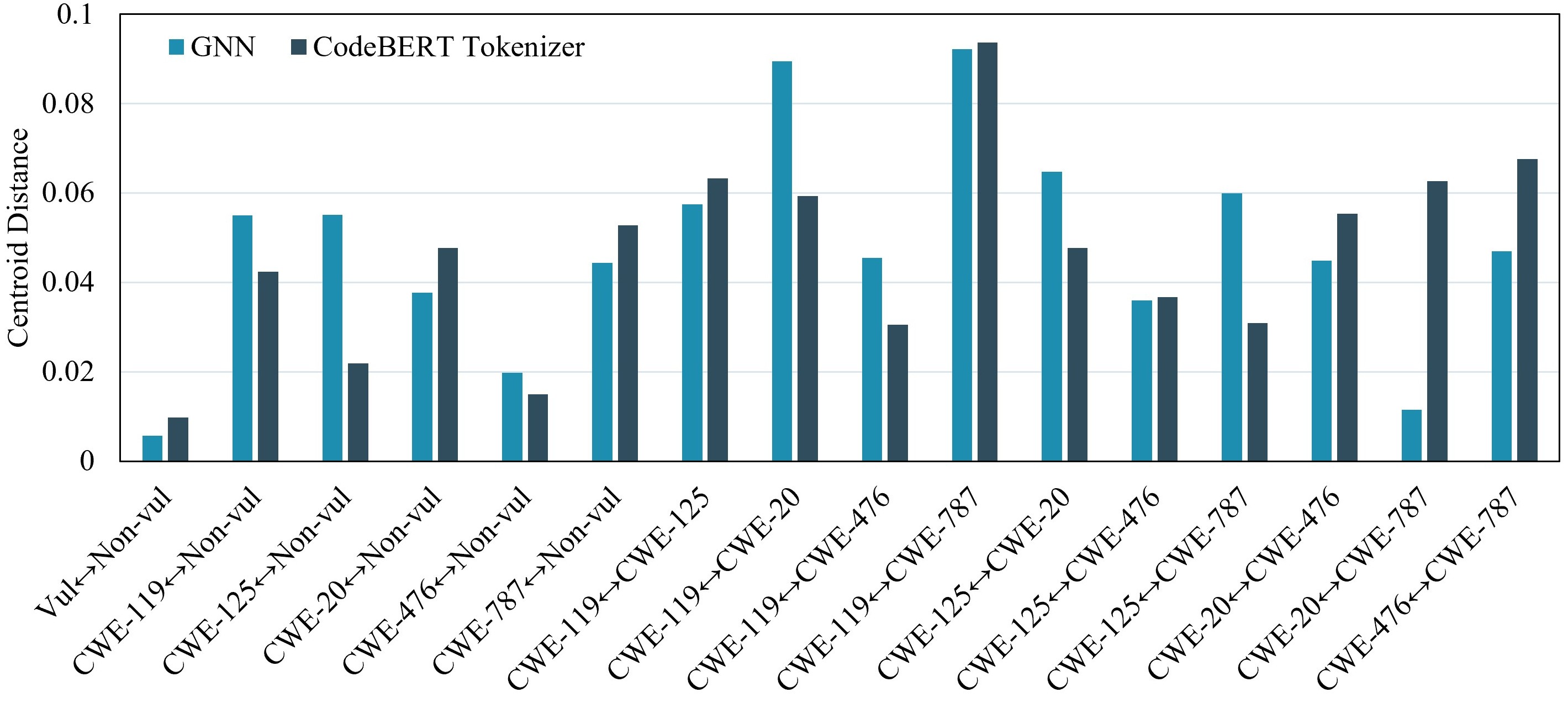}
    \caption{ICVul}
    \label{fig:icvul_cd}
\end{subfigure}
\caption{Centroid distance comparison between top five vulnerability types and non-vulnerable samples in the Juliet, BigVul and ICVul datasets. "Vul" represents the class vulnerable, and "Non-vul" represents the class non-vulnerable.}
\label{fig:centroid_distance_comparison}
\end{figure}

Figure~\ref{fig:centroid_distance_comparison} summarizes the projected centroid distances between the most frequent vulnerability types and non-vulnerable samples. In ICVul (Figure \ref{fig:icvul_cd}), both GNN and CodeBERT show relatively small and erratic centroid distances between different CWE types and non-vulnerable functions. In Juliet (Figure \ref{fig:juliet_cd}), the pattern persists: the distances vary significantly, even within the same dataset, with no consistent dominance of one embedding method over the other. In BigVul (Figure \ref{fig:bigvul_cd}), the centroid distances are generally larger than those in ICVul, especially for CWE-119, CWE-125, and CWE-20. However, this stronger separability is consistent with the earlier visual observation that BigVul may contain inflated inter-class differences due to its labeling strategy. This reveals a key insight—there is no reliable feature-level distinction between certain vulnerability types and non-vulnerable code. Moreover, distances between some vulnerability pairs are even greater than those between vulnerabilities and non-vulnerabilities, indicating that the extracted feature spaces do not exhibit a consistent vulnerability-oriented organization.

Furthermore, we observe that the centroid distances between vulnerability types and non-vulnerable samples in the realistic dataset ICVul are consistently lower than those in Juliet and BigVul. This may be associated with two main factors. First, realistic code tends to be more complex, with higher semantic and structural similarity between vulnerable and non-vulnerable functions, as well as among different vulnerability types, making them harder to distinguish. Second, the datasets contain different CWE distributions and labeling characteristics. Specifically, the top five CWE types in ICVul do not appear in Juliet, while BigVul may introduce larger apparent differences between vulnerable and non-vulnerable samples due to its commit-based labeling strategy.

\begin{tcolorbox}[title=Answer to RQ1]
Before downstream classifier training, graph-based and CodeBERT-based features provide limited and inconsistent separation between vulnerable and non-vulnerable functions across the evaluated datasets. Neither feature extraction method consistently provides vulnerability-oriented separability.
\end{tcolorbox}

\subsection{Benchmark Evaluation (RQ2)}

\subsubsection{Evaluation Results}
Tables~\ref{tab:cross_reveal_results} and~\ref{tab:cross_linevul_results} report the within-dataset and cross-dataset performance of ReVeal, Devign, LineVul, and VulBERTa on Juliet, Devign, BigVul, and ICVul. Cells in the tables are shaded in three levels of gray, where darker shades represent better performance within the same column, helping highlight how well models generalize across datasets when trained on a single source. Bolded values indicate the highest score within each row, showcasing the best model performance on a target dataset when trained on different sources. Each model name in parentheses indicates its training dataset. A, P, R, and F1 denote accuracy, precision, recall, and F1-score, respectively. The diagonal entries correspond to within-dataset evaluation, whereas the off-diagonal entries represent cross-dataset transfer.

\textbf{Within-dataset performance.} When training and testing data originate from the same benchmark, the evaluated models generally achieve their strongest results. This is particularly evident on Juliet, where LineVul and VulBERTa obtain F1-scores of 94.52 and 94.14, respectively. LineVul also reaches an F1-score of 91.00 on BigVul. These results confirm that the models can learn predictive patterns under matched training and testing conditions. However, the variation across models and datasets also shows that within-dataset performance depends strongly on the dataset construction and the compatibility between a model's preprocessing pipeline and the evaluated data.
High within-dataset accuracy does not necessarily indicate effective vulnerability detection. On BigVul, for example, ReVeal, Devign, and VulBERTa obtain accuracies of 87.45, 94.09, and 94.39, respectively, but their corresponding F1-scores are only 23.58, 16.90, and 0.19. BigVul contains a substantially smaller proportion of vulnerable samples than the other benchmarks. Consequently, predictions biased toward the non-vulnerable majority class can produce high accuracy while detecting few vulnerable functions.

\textbf{Cross-dataset performance.} In contrast to the within-dataset results, model performance generally decreases when the training and evaluation data come from different benchmarks. For example, LineVul trained on Juliet achieves an F1-score of 94.52 on Juliet, but its F1-scores decrease to 53.95, 10.73, and 51.22 on Devign, BigVul, and ICVul, respectively. ReVeal trained on BigVul shows an even larger decline, with F1-scores below 1.00 on all three external benchmarks. Similar degradation occurs across the other model--dataset combinations, although its extent varies by model and target dataset. 
This performance degradation should be interpreted in light of the substantial differences among the benchmarks. Juliet contains synthetic and systematically generated samples, Devign consists of manually labeled functions from a limited set of projects, and BigVul and ICVul are constructed from real-world vulnerability-fixing commits using different collection and labeling procedures. The datasets also differ considerably in size and class balance, with vulnerable-sample ratios ranging from 6\% in BigVul to 46\% in Devign. Consequently, a model trained on one benchmark encounters different code characteristics, labeling assumptions, and class distributions when evaluated on another.
No evaluated model consistently preserves its within-dataset performance across the other benchmarks. This indicates that the models learn patterns that are effective under the characteristics of their training data but are not consistently transferable to independently constructed datasets. Therefore, strong within-dataset performance alone does not provide sufficient evidence that a detector has learned robust vulnerability-related patterns.

\begin{table*}[h]
\centering
\footnotesize
\caption{Performance of ReVeal and Devign trained on four benchmark datasets with cross-dataset testing.}
\begin{tabular}{l|cccc|cccc|cccc|cccc}
    \hline
    \textbf{Bench-} &\multicolumn{4}{c|}{\textbf{ReVeal (Juliet)}}&\multicolumn{4}{c|}{\textbf{ReVeal (Devign)}}&\multicolumn{4}{c|}{\textbf{ReVeal (BigVul)}}&\multicolumn{4}{c}{\textbf{ReVeal (ICVul)}}  \\
    \cline{2-17}
     \textbf{mark} & \textbf{\textit{A}}& \textbf{\textit{P}}& \textbf{\textit{R}}& \textbf{\textit{F1}} & \textbf{\textit{A}}& \textbf{\textit{P}}& \textbf{\textit{R}}& \textbf{\textit{F1}} & \textbf{\textit{A}}& \textbf{\textit{P}}& \textbf{\textit{R}}& \textbf{\textit{F1}} & \textbf{\textit{A}}& \textbf{\textit{P}}& \textbf{\textit{R}}& \textbf{\textit{F1}} \\
    \hline
    Juliet & \cellcolor{gray1}{\textbf{\textit{74.87}}} & \cellcolor{gray1}{\textbf{\textit{60.13}}} & \cellcolor{gray1}{\textbf{\textit{92.62}}} & \cellcolor{gray1}{\textbf{\textit{72.92}}} & \cellcolor{gray3}{47.36} & \cellcolor{gray3}{38.84} & \cellcolor{gray3}{76.71} & \cellcolor{gray2}{51.57} & \cellcolor{gray2}{63.38} & \cellcolor{gray3}{32.84} & 0.24 & 0.47 & \cellcolor{gray2}{52.71} & \cellcolor{gray3}{38.27} & \cellcolor{gray3}{48.01} & \cellcolor{gray3}{42.59} \\
    
    Devign & \cellcolor{gray2}{48.59} & \cellcolor{gray2}{46.14} & \cellcolor{gray3}{74.56} & \cellcolor{gray2}{57.01} & \cellcolor{gray1}{\textbf{\textit{49.81}}} & \cellcolor{gray1}{\textbf{\textit{47.59}}} & \cellcolor{gray1}{\textbf{\textit{96.84}}} & \cellcolor{gray1}{\textbf{\textit{63.82}}} & 54.29 & \cellcolor{gray1}{50} & \cellcolor{gray3}{0.25} & \cellcolor{gray3}{0.5} & \cellcolor{gray3}{48.97} & \cellcolor{gray1}{45.86} & \cellcolor{gray2}{64.51} & \cellcolor{gray2}{53.61} \\
    
    BigVul & 19.65 & 5.28 & 74.47 & 9.85 & 30.55 & 6.59 & \cellcolor{gray2}{81.76} & 12.19 & \cellcolor{gray1}{\textbf{\textit{87.45}}} & \cellcolor{gray1}{\textbf{\textit{17.87}}} & \cellcolor{gray1}{\textbf{\textit{34.67}}} & \cellcolor{gray1}{\textbf{\textit{23.58}}} & \cellcolor{gray1}{59.05} & 5.24 & 34.80 & 9.11 \\
    
    ICVul & \cellcolor{gray3}{41.86} & \cellcolor{gray3}{40.23} & \cellcolor{gray2}{86.48} & \cellcolor{gray3}{54.91} & \cellcolor{gray2}{\textbf{\textit{49.08}}} & \cellcolor{gray2}{41.49} & 59.43 & \cellcolor{gray3}{48.86} & \cellcolor{gray3}{58.89} & \cellcolor{gray2}{33.33} & \cellcolor{gray2}{0.41} & \cellcolor{gray2}{0.81} & 42.64 & \cellcolor{gray2}{\textbf{\textit{41.92}}} & \cellcolor{gray1}{\textbf{\textit{95.20}}} & \cellcolor{gray1}{\textbf{\textit{58.21}}} \\
    \hline
    \hline
    \textbf{Bench-} & \multicolumn{4}{c|}{\textbf{Devign (Juliet)}} & \multicolumn{4}{c|}{\textbf{Devign (Devign)}} & \multicolumn{4}{c|}{\textbf{Devign (BigVul)}} & \multicolumn{4}{c}{\textbf{Devign (ICVul)}} \\
    \cline{2-17}
    \textbf{mark} & \textbf{\textit{A}}& \textbf{\textit{P}}& \textbf{\textit{R}}& \textbf{\textit{F1}} & \textbf{\textit{A}}& \textbf{\textit{P}}& \textbf{\textit{R}}& \textbf{\textit{F1}} & \textbf{\textit{A}}& \textbf{\textit{P}}& \textbf{\textit{R}}& \textbf{\textit{F1}} & \textbf{\textit{A}}& \textbf{\textit{P}}& \textbf{\textit{R}}& \textbf{\textit{F1}} \\
    \hline
    Juliet & \cellcolor{gray2}{\textbf{\textit{79.45}}} & \cellcolor{gray1}{\textbf{\textit{74.77}}} & \cellcolor{gray1}{\textbf{\textit{66.04}}} & \cellcolor{gray1}{\textbf{\textit{70.13}}} & \cellcolor{gray1}{58.16} & \cellcolor{gray3}{34.34} & 15.94 & \cellcolor{gray3}{21.77} & \cellcolor{gray2}{63.47} & 1.00 & 0.01 & 0.02 & \cellcolor{gray1}{57.53} & \cellcolor{gray2}{36.41} & 21.79 & \cellcolor{gray3}{27.26} \\

    Devign & \cellcolor{gray3}{50.06} & \cellcolor{gray2}{44.15} & \cellcolor{gray3}{34.80} & \cellcolor{gray3}{38.92} & \cellcolor{gray2}{\textbf{\textit{55.22}}} & \cellcolor{gray1}{\textbf{\textit{50.72}}} & \cellcolor{gray1}{\textbf{\textit{73.59}}} & \cellcolor{gray1}{\textbf{\textit{60.05}}} & 54.35 & \cellcolor{gray1}{56.25} & \cellcolor{gray2}{0.75} & \cellcolor{gray2}{1.48} & \cellcolor{gray2}{52.34} & \cellcolor{gray1}{46.56} & \cellcolor{gray3}{28.65} & \cellcolor{gray1}{35.48} \\

    BigVul & \cellcolor{gray1}{80.10} & 6.99 & 19.30 & 10.26 & \cellcolor{gray3}{54.33} & 4.02 & \cellcolor{gray3}{29.48} & 7.08 & \cellcolor{gray1}{94.09} & \cellcolor{gray2}{49.63} & \cellcolor{gray1}{10.18} & \cellcolor{gray1}{16.90} & \cellcolor{gray3}{49.61} & 6.73 & \cellcolor{gray1}{58.66} & 12.07 \\

    ICVul & 49.96 & \cellcolor{gray3}{41.14} & \cellcolor{gray2}{51.84} & \cellcolor{gray2}{45.87} & 46.27 & \cellcolor{gray2}{40.99} & \cellcolor{gray2}{\textbf{\textit{71.31}}} & \cellcolor{gray2}{\textbf{\textit{52.06}}} & \cellcolor{gray3}{58.93} & \cellcolor{gray3}{37.50} & \cellcolor{gray3}{0.61} & \cellcolor{gray3}{1.21} & 42.83 & \cellcolor{gray3}{31.35} & \cellcolor{gray2}{33.40} & \cellcolor{gray2}{32.34} \\
    \hline
\end{tabular}
\label{tab:cross_reveal_results}
\end{table*}

\begin{table*}
\centering
\footnotesize
\caption{Performance of LineVul and VulBERTa trained on four benchmark datasets with cross-dataset testing.}
\begin{tabular}{l|cccc|cccc|cccc|cccc}
    \hline
    \textbf{Bench-} &\multicolumn{4}{c|}{\textbf{LineVul (Juliet)}}&\multicolumn{4}{c|}{\textbf{LineVul (Devign)}}&\multicolumn{4}{c|}{\textbf{LineVul (BigVul)}}&\multicolumn{4}{c}{\textbf{LineVul (ICVul)}}  \\
    \cline{2-17}
     \textbf{mark} & \textbf{\textit{A}}& \textbf{\textit{P}}& \textbf{\textit{R}}& \textbf{\textit{F1}} & \textbf{\textit{A}}& \textbf{\textit{P}}& \textbf{\textit{R}}& \textbf{\textit{F1}} & \textbf{\textit{A}}& \textbf{\textit{P}}& \textbf{\textit{R}}& \textbf{\textit{F1}} & \textbf{\textit{A}}& \textbf{\textit{P}}& \textbf{\textit{R}}& \textbf{\textit{F1}} \\
    \hline
    Juliet & \cellcolor{gray1}{\textbf{\textit{95.98}}} & \cellcolor{gray1}{\textbf{\textit{95.16}}} & \cellcolor{gray1}{93.90} & \cellcolor{gray1}{\textbf{\textit{94.52}}} & \cellcolor{gray3}{38.20} & \cellcolor{gray3}{36.92} & \cellcolor{gray2}{\textbf{\textit{94.71}}} & \cellcolor{gray3}{53.13} & \cellcolor{gray2}{62.97} & 0 & 0 & 0 & \cellcolor{gray1}{63.86} & \cellcolor{gray1}{52.68} & 22.27 & \cellcolor{gray3}{31.30} \\
    
    Devign & \cellcolor{gray2}{49.01} & \cellcolor{gray2}{45.79} & 65.65 & \cellcolor{gray2}{53.95} & \cellcolor{gray1}{\textbf{\textit{66.54}}} & \cellcolor{gray1}{\textbf{\textit{65.68}}} & 55.43 & \cellcolor{gray1}{\textbf{\textit{60.12}}} & 54.69 & \cellcolor{gray2}{63.16} & \cellcolor{gray3}{0.97} & \cellcolor{gray3}{1.90} & \cellcolor{gray2}{47.04} & \cellcolor{gray2}{45.42} & \cellcolor{gray1}{\textbf{\textit{81.42}}} & \cellcolor{gray1}{58.31} \\
    
    BigVul & 27.57 & 5.76 & \cellcolor{gray2}{77.60} & 10.73 & 11.43 & 4.96 & \cellcolor{gray3}{81.38} & 9.34 & \cellcolor{gray1}{\textbf{\textit{96}}} & \cellcolor{gray1}{\textbf{\textit{97}}} & \cellcolor{gray1}{\textbf{\textit{86}}} & \cellcolor{gray1}{\textbf{\textit{91}}} & 39.33 & 6.42 & \cellcolor{gray3}{72.31} & 11.79\\
    
    ICVul & \cellcolor{gray3}{44.09} & \cellcolor{gray3}{38.77} & \cellcolor{gray3}{75.46} & \cellcolor{gray3}{51.22} & \cellcolor{gray2}{39.16} & \cellcolor{gray2}{38.93} & \cellcolor{gray1}{\textbf{\textit{99.17}}} & \cellcolor{gray2}{\textbf{\textit{55.91}}} & \cellcolor{gray3}{57.6} & \cellcolor{gray3}{32.24} & \cellcolor{gray2}{8.18} & \cellcolor{gray2}{13.05} & \cellcolor{gray3}{\textbf{\textit{44.35}}} & \cellcolor{gray3}{\textbf{\textit{39.01}}} & \cellcolor{gray2}{76.46} & \cellcolor{gray2}{51.66} \\
    \hline
    \hline
    \textbf{Bench-} & \multicolumn{4}{c|}{\textbf{VulBERTa (Juliet)}} & \multicolumn{4}{c|}{\textbf{VulBERTa (Devign)}} & \multicolumn{4}{c|}{\textbf{VulBERTa (BigVul)}} & \multicolumn{4}{c}{\textbf{VulBERTa (ICVul)}} \\
    \cline{2-17}
    \textbf{mark} & \textbf{\textit{A}}& \textbf{\textit{P}}& \textbf{\textit{R}}& \textbf{\textit{F1}} & \textbf{\textit{A}}& \textbf{\textit{P}}& \textbf{\textit{R}}& \textbf{\textit{F1}} & \textbf{\textit{A}}& \textbf{\textit{P}}& \textbf{\textit{R}}& \textbf{\textit{F1}} & \textbf{\textit{A}}& \textbf{\textit{P}}& \textbf{\textit{R}}& \textbf{\textit{F1}} \\
    \hline
    Juliet & \cellcolor{gray1}{\textbf{\textit{95.40}}} & \cellcolor{gray1}{\textbf{\textit{89.05}}} & \cellcolor{gray1}{\textbf{\textit{99.84}}} & \cellcolor{gray1}{\textbf{\textit{94.14}}} & \cellcolor{gray2}{56.71} & \cellcolor{gray2}{41.80} & 43.55 & \cellcolor{gray3}{42.66} & \cellcolor{gray2}{63.03} & 0 & 0 & 0 & \cellcolor{gray3}{38.93} & \cellcolor{gray3}{37.56} & \cellcolor{gray1}{98.34} & \cellcolor{gray3}{54.35} \\

    Devign & \cellcolor{gray2}{45.49} & \cellcolor{gray2}{44.91} & 87.37 & \cellcolor{gray2}{59.33} & \cellcolor{gray1}{\textbf{\textit{62.55}}} & \cellcolor{gray1}{\textbf{\textit{56.49}}} & \cellcolor{gray2}{77.07} & \cellcolor{gray1}{\textbf{\textit{65.19}}} & 54.50 & 0 & 0 & 0 & \cellcolor{gray1}{46.49} & \cellcolor{gray1}{45.69} & \cellcolor{gray3}{\textbf{\textit{93.48}}} & \cellcolor{gray1}{61.38} \\

    BigVul & 9.53 & 5.63 & \cellcolor{gray2}{95.94} & 10.64 & 37.87 & 6.37 & \cellcolor{gray3}{73.44} & 11.72 & \cellcolor{gray1}{94.39} & \cellcolor{gray1}{50.00} & \cellcolor{gray1}{0.09} & \cellcolor{gray1}{0.19} & 9.53 & 5.63 & \cellcolor{gray2}{95.94} & 10.64 \\

    ICVul & \cellcolor{gray3}{38.64} & \cellcolor{gray3}{38.28} & \cellcolor{gray3}{\textbf{\textit{94.32}}} & \cellcolor{gray3}{54.46} & \cellcolor{gray3}{\textbf{\textit{45.52}}} & \cellcolor{gray3}{39.86} & \cellcolor{gray1}{78.80} & \cellcolor{gray2}{52.94} & \cellcolor{gray3}{61.10} & 0 & 0 & 0 & \cellcolor{gray2}{45.13} & \cellcolor{gray2}{\textbf{\textit{40.49}}} & 87.48 & \cellcolor{gray2}{\textbf{\textit{55.36}}} \\
    \hline
\end{tabular}
\label{tab:cross_linevul_results}
\end{table*}

The observed cross-dataset behavior is associated with several dataset characteristics, including labeling strategy, class balance, dataset size, data origin, and project scope. BigVul combines patch-based labeling with substantial class imbalance, and models trained on it often exhibit high accuracy but low recall or F1-score outside their matched evaluation setting. In contrast, models trained on Juliet, Devign, or ICVul exhibit different transfer patterns, although no training dataset produces consistently strong results across all target benchmarks. Juliet provides a large set of systematically generated samples but differs considerably from real-world code. Devign contains manually labeled functions from a limited set of projects, while ICVul uses a patch-derived collection and filtering procedure intended to reduce labeling uncertainty.
Because these dataset characteristics vary simultaneously, the current experiments cannot isolate the effect of any individual factor.
The results instead show that model performance is sensitive to the combined characteristics of the source and target datasets.

Model architecture alone does not explain the observed transfer behavior. Both graph-based and transformer-based models degrade under cross-dataset evaluation, although the extent and direction of degradation vary across model--dataset combinations. Model-specific preprocessing, class distribution, data origin, project scope, and labeling procedures may all affect the patterns available for learning. The current results do not isolate the contribution of each factor, but they consistently show that strong performance under matched benchmark conditions does not transfer reliably to independently constructed datasets.

\subsubsection{Failure Modes}
\textbf{Failures associated with BigVul preprocessing and class imbalance.} VulBERTa, Devign, and ReVeal trained on BigVul obtain high or moderate within-dataset accuracy but very low recall and F1-scores. VulBERTa (BigVul) achieves an F1-score of only 0.19 on BigVul and zero on all external benchmarks, while Devign (BigVul) and ReVeal (BigVul) also detect few vulnerable samples. For VulBERTa and Devign, the model-specific preprocessing pipelines filter out part of BigVul before training, reducing the effective training data. This data loss is particularly consequential because only 6\% of BigVul samples are vulnerable. Although ReVeal applies SMOTE, resampling cannot improve the quality or diversity of the original vulnerable samples or remove possible noise introduced by commit-based labeling. Thus, preprocessing-induced data reduction, severe class imbalance, and labeling limitations collectively restrict the vulnerability-related patterns available for learning.

\textbf{Benchmark-specific learning.} LineVul (BigVul) presents a different failure pattern. It achieves an F1-score of 91.00 within BigVul, but its F1-score decreases to zero on Juliet, 1.90 on Devign, and 13.05 on ICVul. Its failure therefore cannot be explained solely by insufficient within-dataset learning. Instead, the large performance gap suggests that the lexical and contextual patterns learned from BigVul do not remain predictive when evaluated on datasets constructed from different data sources and labeling procedures.
Similar degradation occurs for models trained on Juliet, Devign, and ICVul. Juliet provides a large volume of consistently labeled synthetic samples, but its controlled code patterns may not transfer effectively to real-world functions. Devign contains manually labeled real-world functions but covers a limited number of projects, while the smaller size of ICVul may restrict the diversity of training samples. These differences in data origin, size, project scope, class balance, and labeling strategy affect the patterns learned by the models and help explain why strong performance on one benchmark is not consistently reproduced on another.

Overall, the failures occur across both graph-based and transformer-based architectures and cannot be attributed to a single model family. Model-specific preprocessing can reduce or alter the effective training data, while class imbalance, dataset size, data origin, project scope, and labeling strategy influence which patterns are learned. Cross-dataset evaluation therefore reveals model and dataset dependencies that are not visible from within-dataset results alone.

\begin{tcolorbox}[title=Answer to RQ2]
Models generally perform best under within-dataset evaluation but degrade substantially and inconsistently across independently constructed benchmarks. Thus, strong within-dataset performance does not necessarily indicate transferable vulnerability detection.
\end{tcolorbox}

\subsection{Deployment Evaluation (RQ3)}

\subsubsection{DL Models}

As shown in Table~\ref{tab:dl_ventivul}, all trained DL models exhibit limited performance on the temporally separated VentiVul benchmark. Their F1-scores range from 3.47 to 8.65, despite substantial differences in accuracy and recall. Several configurations show strongly biased prediction behavior. ReVeal and LineVul trained on Devign reach recall values close to or equal to 100\%, but their precision remains approximately 2\%, indicating extensive false-positive predictions. In contrast, the BigVul-trained models obtain higher accuracy but substantially lower recall. ReVeal (BigVul) achieves the highest DL F1-score of 8.65, but its precision is only 5.25\%. These results show that none of the trained models provides a practically reliable balance between detecting recent vulnerable functions and avoiding false alarms.

The Function-Pair results reveal a further limitation. Several models achieve high CDR because they correctly detect at least one vulnerable function for many CVEs. For example, ReVeal (Devign), ReVeal (ICVul), and LineVul (Devign) obtain CDR values of 99.10\%, 94.59\%, and 100\%, respectively. However, their corresponding PACC values are only 0.76\%, 0\%, and 0\%. The large difference between CDR and PACC shows that detecting a vulnerable function does not imply that a model can distinguish it from its patched counterpart. Across all DL configurations, PACC remains below 5.5\%, with ReVeal (BigVul) and LineVul (ICVul) obtaining the highest values of 5.34\% and 5.32\%, respectively. Therefore, the trained DL models show limited sensitivity to the security-relevant code changes introduced by vulnerability patches.

\begin{table}[h]
\centering
\small
\caption{Performance of ReVeal (RV) and LineVul (LV) on VentiVul under Whole-File and Function-Pair settings.}
\label{tab:dl_ventivul}
\begin{tabular}{l|cccc|cc}
\hline
\multirow{2}{*}{Model} & \multicolumn{4}{c|}{Whole-File} & \multicolumn{2}{c}{Function-Pair} \\
\cline{2-7}
& A & P & R & F1 & CDR & PACC \\
\hline
RV (Juliet) & 46.5  & 2.05 & 61.83 & 3.97 & 63.06 & 4.58 \\
RV (Devign) & 11.61 & 1.97 & 99.24 & 3.86 & 99.1  & 0.76 \\
RV (BigVul) & 90.77 & 5.25 & 24.43 & 8.65 & 27.03 & 5.34 \\
RV (ICVul)  & 6.51  & 1.77 & 93.89 & 3.47 & 94.59 & 0 \\
\hline
LV (Juliet) & 30.01 & 2.25 & 81.37 & 4.39 & 82    & 1.52 \\
LV (Devign) & 3.38  & 2    & 100   & 3.93 & 100   & 0 \\
LV (BigVul) & 81.28 & 3.81 & 34.98 & 6.87 & 38    & 2.28 \\
LV (ICVul)  & 42.47 & 2.21 & 65.02 & 4.27 & 73    & 5.32 \\
\hline
\end{tabular}
\end{table}

\subsubsection{LLMs}

\begin{table*}[h!]
\centering
\small
\caption{Performance of LLMs on VentiVul under different prompting strategies. Each cell reports Conservative/Valid-only results. All values are percentages.}
\label{tab:llm_ventivul}
\begin{tabular}{l|ccc|ccc|ccc}
\hline
\multirow{2}{*}{\textbf{Model}}
& \multicolumn{3}{c|}{\textbf{Zero-shot}}
& \multicolumn{3}{c|}{\textbf{Few-shot}}
& \multicolumn{3}{c}{\textbf{CoT}} \\
\cline{2-10}
& \textbf{F1} & \textbf{CDR} & \textbf{PACC}
& \textbf{F1} & \textbf{CDR} & \textbf{PACC}
& \textbf{F1} & \textbf{CDR} & \textbf{PACC} \\
\hline
Qwen2.5-Coder-32B
& 3.72/5.51 & 9/9 & 3.04/3.05
& 2.95/3.88 & 7.5/7.5 & 2.68/2.68
& 3.04/8.21 & 14.5/15.51 & 1.52/1.66 \\

DeepSeek-Coder-V2-Lite
& 4.29/4.32 & 34/34 & 5.7/5.73
& 3.64/3.66 & 7.5/7.5 & 1.53/1.53
& 4.89/4.89 & 23.5/23.5 & 4.56/4.56 \\

Qwen3-30B-A3B
& 3.09/3.21 & 5/5 & 1.14/1.14
& 0.79/1.45 & 1/1.03 & 0.38/0.4
& 1.39/5.67 & 7.5/9.15 & 1.9/2.53 \\

Mistral-Small-24B
& 4.07/4.07 & 20/20 & 4.18/4.18
& 5.36/5.37 & 46/46 & 5.75/5.75
& 4.71/4.71 & 8.5/8.5 & 1.52/1.52 \\
\hline
\end{tabular}
\end{table*}

Table~\ref{tab:llm_ventivul} reports the performance of the four off-the-shelf LLMs under zero-shot, few-shot, and CoT prompting. Overall, the LLMs also exhibit weak deployment-oriented vulnerability detection. All conservative F1-scores remain below 5.5\%, and all PACC values remain below 6\%. Mistral-Small-24B with few-shot prompting achieves the highest conservative F1-score, CDR, and PACC, reaching 5.36\%, 46\%, and 5.75\%, respectively. However, even this configuration correctly differentiates only a small proportion of vulnerable--patched pairs. Prompting does not provide consistent improvements. Few-shot prompting improves all three reported metrics for Mistral-Small-24B but degrades performance for the other models. CoT increases the valid-only F1-score of Qwen2.5-Coder-32B, but its PACC decreases, showing that improved function-level classification does not necessarily produce stronger sensitivity to vulnerability-fixing changes. The effectiveness of prompting is therefore model- and metric-dependent rather than generally beneficial.

The conservative and valid-only policies further expose the practical impact of malformed or unsupported outputs. Their differences are most visible for Qwen2.5-Coder-32B and Qwen3-30B-A3B under CoT, whereas DeepSeek-Coder-V2-Lite and Mistral-Small-24B produce nearly identical results under the two policies. Nevertheless, valid formatting alone does not imply effective vulnerability detection: models with more consistently parseable responses still obtain low F1 and PACC. Overall, the evaluated LLMs do not demonstrate reliable detection or patch-pair differentiation without task-specific adaptation.

\subsubsection{Project-specific Comparison}
\begin{figure*}[h]
    \centering
    \includegraphics[width=1\textwidth]{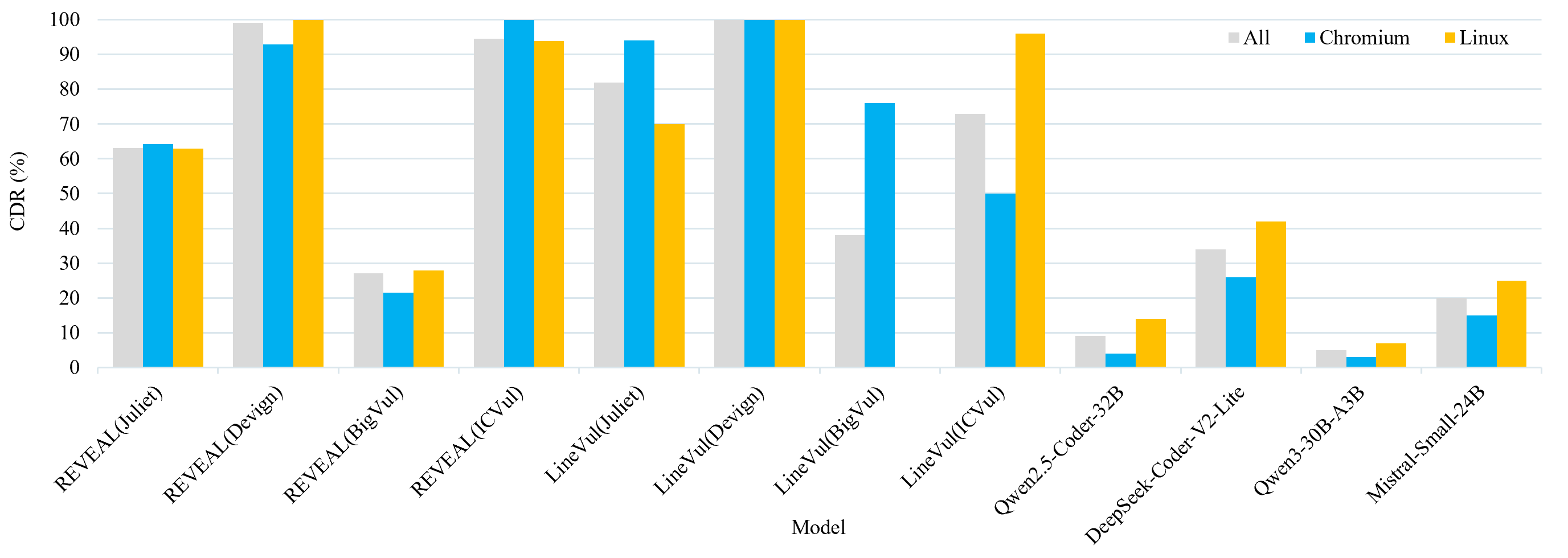}
    \caption{Project-specific CDR of the evaluated DL models and LLMs on the Chromium and Linux subsets of VentiVul. For the LLMs, zero-shot results are reported. ``All'' denotes performance on the complete benchmark.}
    \label{fig:project_cacc}
\end{figure*}

\begin{figure*}[h]
    \centering
    \includegraphics[width=1\textwidth]{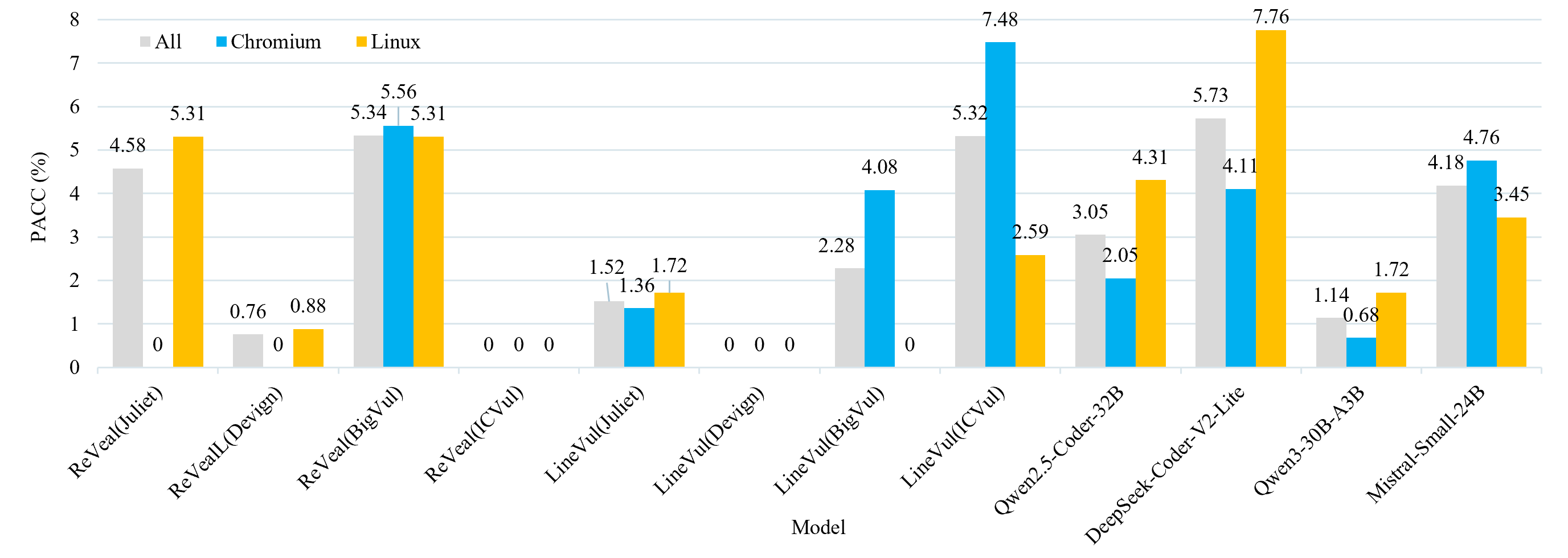}
    \caption{Project-specific PACC of the evaluated DL models and LLMs on the Chromium and Linux subsets of VentiVul. For the LLMs, zero-shot results are reported. ``All'' denotes performance on the complete benchmark.}
    \label{fig:project_pacc}
\end{figure*}

Figures~\ref{fig:project_cacc} and~\ref{fig:project_pacc} compare CDR and PACC on the Chromium and Linux subsets of VentiVul. For each DL architecture, the training dataset is shown in parentheses. 
For the LLMs, we report the valid-only results under zero-shot prompting as a representative configuration. Zero-shot prompting provides a common evaluation setting across all LLMs without introducing variation from demonstration selection or prompt-specific examples. In addition, under zero-shot prompting, the differences between the valid-only and conservative PACC results are negligible. This selection supports a concise project-specific comparison without presenting every model--prompt--policy combination.

The project-specific CDR results vary substantially across models. Several DL configurations detect at least one vulnerable function for most CVEs. LineVul (Devign) obtains 100\% CDR on both Chromium and Linux. ReVeal (Devign) reaches 92.86\% on Chromium and 100\% on Linux, while ReVeal (ICVul) reaches 100\% and 93.81\%, respectively. However, these configurations also exhibit recall close to or equal to 100\% and precision of approximately 2\% in the Whole-File evaluation. Their high CDR therefore reflects broad vulnerability coverage but does not indicate precise detection. Other DL configurations exhibit substantial differences between the two projects. LineVul (BigVul) obtains 76\% CDR on Chromium but fails to detect any Linux CVE, whereas LineVul (ICVul) obtains 50\% on Chromium and 96\% on Linux. ReVeal (Juliet) is more consistent across the projects, reaching 64.29\% on Chromium and 62.89\% on Linux, but its CDR remains lower than that of the strongly positive-biased configurations. These results show that CVE coverage depends on both the source training dataset and the target project, with no training dataset producing consistently strong behavior across both DL architectures.
Under zero-shot prompting, the LLMs obtain lower CDR than most DL configurations. DeepSeek-Coder-V2-Lite achieves the highest overall LLM CDR at 34\%, comprising 26\% on Chromium and 42\% on Linux. Mistral-Small-24B follows with 20\% overall, while Qwen2.5-Coder-32B and Qwen3-30B-A3B obtain only 9\% and 5\%, respectively. All four LLMs achieve higher CDR on Linux than on Chromium, although their absolute CVE detection rates remain limited.

The project-specific PACC results provide a substantially stricter view. Despite the high CDR of several DL configurations, the best project-specific PACC remains below 8\%. LineVul (ICVul) achieves the highest DL PACC on Chromium at 7.48\%, but decreases to 2.59\% on Linux. ReVeal (Juliet) obtains 5.31\% on Linux but fails to differentiate any patch pair on Chromium. ReVeal (BigVul) shows more similar results across the two projects, reaching 5.56\% on Chromium and 5.31\% on Linux. Among the LLMs, DeepSeek-Coder-V2-Lite obtains 4.11\% on Chromium and 7.76\% on Linux, while the other LLMs also remain low and project-dependent. Consequently, neither high CDR nor relative stability across projects translates into reliable patch-pair differentiation.

\subsubsection{CWE-specific Comparison}
\begin{figure*}[h]
    \centering
    \includegraphics[width=1\textwidth]{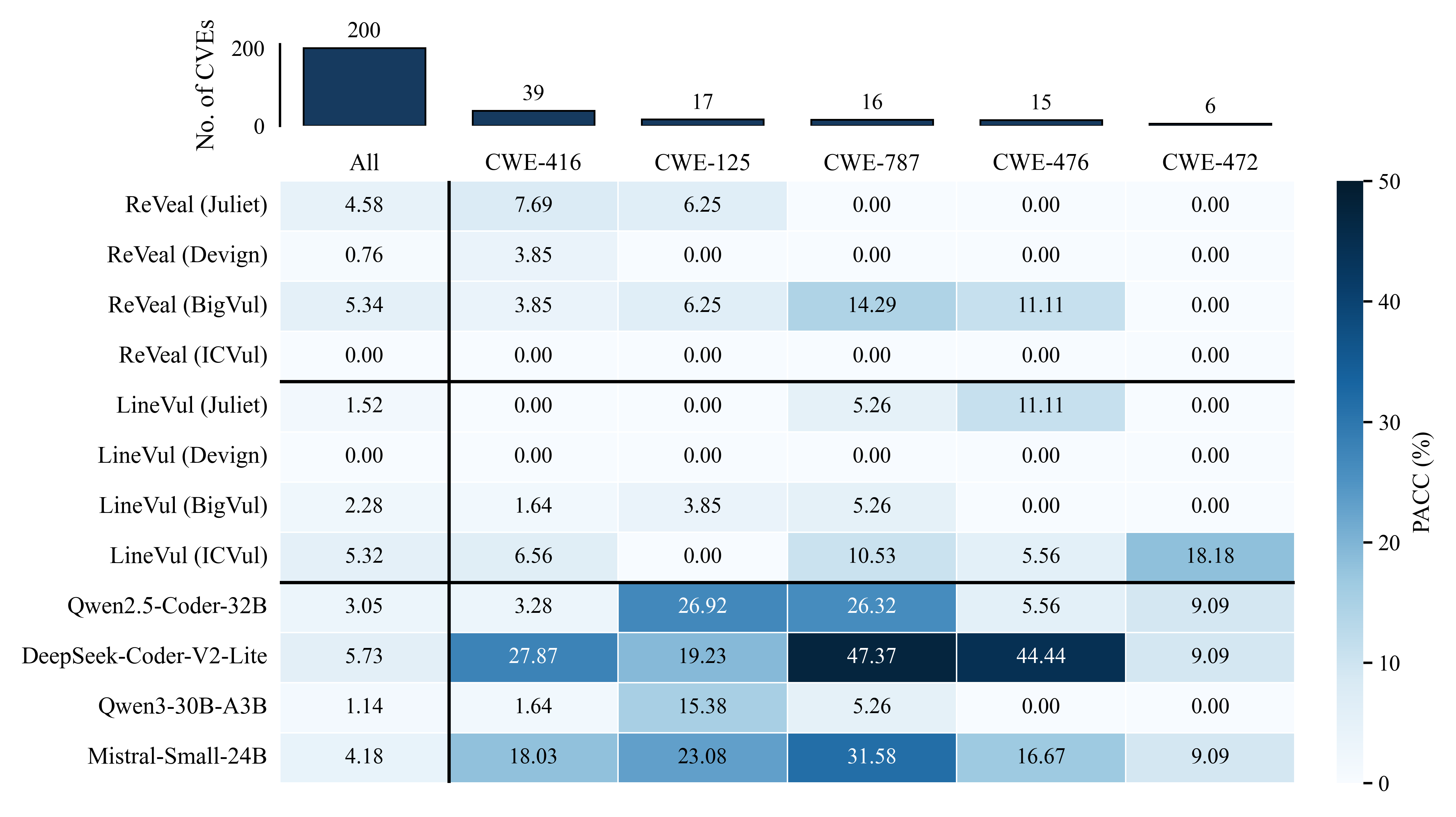}
    \caption{CWE-specific PACC of the evaluated DL models and LLMs on VentiVul. The upper panel shows the number of CVEs in the complete benchmark and in each selected CWE category. For the LLMs, zero-shot results are reported. ``All'' denotes performance on the complete benchmark.}
    \label{fig:cwe_pacc}
\end{figure*}

Figure~\ref{fig:cwe_pacc} compares PACC across the five most frequent CWE categories in VentiVul. The results vary considerably across vulnerability types. Among the DL configurations, ReVeal (BigVul) obtains relatively higher PACC for CWE-787 and CWE-476, reaching 14.29\% and 11.11\%, respectively. LineVul (ICVul) reaches 18.18\% for CWE-472 and 10.53\% for CWE-787. However, these improvements are limited to individual categories, and several DL configurations obtain zero PACC across all five selected CWEs. Differences among configurations using the same architecture further indicate that the source training dataset affects which vulnerability types can be differentiated.
The LLMs obtain relatively higher PACC for some individual CWE categories despite their low aggregate performance. DeepSeek-Coder-V2-Lite achieves the strongest results for several categories, including CWE-787 and CWE-476, while Qwen2.5-Coder-32B performs best for CWE-125. Nevertheless, no LLM performs consistently well across all selected categories. Because the CWE categories contain different numbers of CVEs, these percentages should be interpreted together with the sample counts shown in the figure. The relatively high values for individual model--CWE combinations therefore represent localized successes rather than stable generalization across vulnerability types. Overall, patch-pair differentiation remains strongly dependent on the vulnerability category, model, and training configuration.

\begin{tcolorbox}[title=Answer to RQ3]
Both trained DL models and off-the-shelf LLMs show limited effectiveness on recent, temporally separated vulnerabilities. Although some configurations detect vulnerable functions or CVEs, their low PACC demonstrates that they rarely distinguish vulnerable code reliably from its patched counterpart.
\end{tcolorbox}

\section{Discussion}

Our evaluation reveals a consistent gap between performance under controlled benchmark conditions and effectiveness under distribution shift and deployment-oriented evaluation. Across the three RQs, the results suggest that this gap cannot be attributed to a single architecture, dataset, or evaluation metric. Instead, model behavior is jointly shaped by the adopted representations, benchmark construction, model-specific data processing, and the conditions under which performance is measured. This section discusses the implications of these findings for vulnerability representation, benchmark design, model evaluation, and future research.

\subsection{Pre-training Feature Representations Provide Limited Separability}
RQ1 evaluates the feature spaces produced by the graph-based and CodeBERT-based extraction pipelines before downstream classifier training. The analysis therefore concerns the effectiveness of the feature representations themselves, rather than representations learned or reshaped through supervised vulnerability classification. Neither feature extraction method provides consistently separable vulnerable and non-vulnerable functions across the four benchmarks. CodeBERT features show relatively stronger separation for some datasets, while graph-based features show comparable or stronger separation in other cases. The results therefore do not identify a representation method that consistently exposes discriminative vulnerability-related signals across heterogeneous datasets.

The observed separability also varies across datasets and CWE categories, indicating that the extracted feature spaces may reflect dataset-specific code and labeling characteristics rather than a stable vulnerability-oriented organization. In realistic code, vulnerable and non-vulnerable functions may differ through small security-relevant operations or conditions while retaining similar syntax and structure, making such distinctions difficult to expose through general-purpose structural or token-based representations.

These observations must nevertheless be interpreted cautiously. The t-SNE plots are low-dimensional projections, and centroid distances provide descriptive measurements of the extracted feature spaces. They do not establish what a subsequently trained classifier can or cannot learn, nor do they prove that limited initial separability directly causes poor classification performance. Instead, RQ1 shows that the two input feature extraction methods do not provide an immediately clear and consistent vulnerability-oriented structure before classifier training. Downstream models must therefore learn their decision boundaries from feature spaces in which vulnerability-related distinctions are limited, inconsistent, or entangled with other code characteristics.

\subsection{Benchmark Performance Depends on Model and Dataset Factors}
RQ2 addresses a different question: whether supervised DL models trained on these benchmark datasets maintain their performance when the evaluation dataset changes. The models generally perform best under within-dataset evaluation but degrade substantially and inconsistently under cross-dataset testing. This shows that successful classifier training under matched conditions does not ensure robustness across independently constructed benchmarks. The RQ2 results should not be treated as a direct consequence of the feature separability observed in RQ1. Supervised training can reshape or combine the available features, and the evaluated architectures use different learning mechanisms. Nevertheless, the two RQs provide complementary evidence. RQ1 shows that the initial graph-based and token-based features do not consistently expose clear vulnerability-related distinctions, while RQ2 shows that downstream training does not produce models whose performance remains stable across datasets.

Cross-dataset failures reflect interactions between model and dataset factors. Differences in data origin, project scope, class balance, labeling strategy, and model-specific preprocessing jointly affect the effective data available for learning. Consequently, architecture comparisons should consider not only the original benchmark statistics but also the samples retained after preprocessing. Because these characteristics vary simultaneously, the observed degradation should be interpreted as sensitivity to changing benchmark conditions rather than as evidence that any single dataset characteristic independently determines generalization.

\subsection{Deployment Evaluation Reveals a Stricter Capability Gap}
RQ3 evaluates trained DL models and off-the-shelf LLMs on recent, temporally separated vulnerabilities. Both model types exhibit limited function-level performance on VentiVul. Some configurations achieve high recall or CDR, but these results are often accompanied by low precision, showing that broad vulnerability coverage may result from excessive positive predictions. The gap between CDR and PACC provides the clearest deployment-oriented finding. The consistently low PACC shows that detecting some vulnerable functions does not imply sensitivity to the security-relevant changes introduced by their patches.

Project- and CWE-specific results reveal localized differences, but no configuration performs consistently across projects and vulnerability categories. Likewise, zero-shot, few-shot, and CoT prompting produce model-dependent effects for the LLMs, rather than consistent improvements. Invalid responses introduce an additional practical limitation, although valid output formatting alone does not ensure correct vulnerability detection. RQ3 therefore evaluates a capability not captured by conventional benchmark testing: whether a detector can identify recent vulnerabilities and respond correctly after the vulnerable condition has been removed. The results show that current trained DL models and off-the-shelf LLMs remain unreliable under this stricter setting.

\subsection{Implications and Future Directions}
The combined findings suggest three priorities. First, feature representation and model training should more directly emphasize security-relevant code changes. Vulnerable--patched pairs may provide useful supervision for distinguishing the small semantic differences that alter a function's security status. Second, dataset construction should prioritize reliable labels, diverse vulnerability patterns, and explicit links between vulnerable code and its corresponding fixes, rather than focusing primarily on dataset size. Third, within-dataset evaluation should be complemented by cross-dataset and temporally separated testing. Function-level metrics should be reported together with CVE-level coverage and patch-pair differentiation because these settings evaluate distinct capabilities. Future work may further examine individual dataset factors, representation semantics, broader software ecosystems, and more advanced LLM adaptation strategies.

Overall, the results identify limitations at two related but distinct levels. The evaluated pre-training feature extraction methods do not consistently provide separable vulnerability-related signals, and the trained detectors do not maintain reliable performance across datasets or under temporally separated deployment evaluation. Addressing both limitations is necessary for improving the transferability and practical reliability of vulnerability detection systems.

\section{Threats and Limitations}
Our study is subject to several limitations concerning model coverage, representation analysis, dataset generalizability, and VentiVul construction.

First, our benchmark evaluation covers four representative DL models, while the deployment evaluation focuses on two DL model families and four open-weight LLMs. These models represent both graph-based and transformer-based approaches, but they do not cover all recently proposed vulnerability detectors or proprietary LLMs. The conclusions should therefore be interpreted as applying to the evaluated models and settings rather than to all existing vulnerability detection techniques.
The model coverage was constrained by the requirements of reproducibility, experimental consistency, and computational feasibility.
Moreover, because the complete pretraining corpora and knowledge cutoffs of the evaluated LLMs are unavailable, temporal separation based on public release and disclosure dates cannot fully exclude training-data contamination.

Second, RQ1 analyzes the graph-based and CodeBERT feature spaces before downstream classifier training. The t-SNE visualizations and centroid distances provide complementary views of whether the extracted features contain readily separable vulnerability-related signals. However, t-SNE is sensitive to dimensionality-reduction parameters, and centroid distance cannot fully characterize complex or nonlinear class structures. These analyses therefore do not determine what a subsequently trained classifier can learn, nor do they establish a causal relationship between initial feature separability and the performance observed in RQ2 and RQ3.

Third, the evaluated datasets differ in size, class balance, project scope, data origin, and labeling strategy. Because these characteristics vary simultaneously, the cross-dataset experiments reveal sensitivity to changing benchmark conditions but cannot isolate the causal effect of each individual characteristic. Some model-specific preprocessing pipelines exclude functions that cannot be converted into the required inputs, which may affect the comparability of individual model--dataset configurations. Moreover, ICVul and VentiVul are derived from open-source projects and may not fully represent the coding practices, architectures, or vulnerability distributions of proprietary industrial software.

Finally, VentiVul is limited to C/C++ code from Linux and Chromium and does not contain every vulnerability reported for the selected project versions. Its results may therefore not generalize to other programming languages, smaller projects, or proprietary software.
Under the Whole-File setting, functions unrelated to the collected CVEs are treated as non-vulnerable. However, an apparently unrelated or patched function may be associated with another vulnerability not included in VentiVul or discovered later. Consequently, the presumed non-vulnerable labels in the Whole-File setting may contain residual label noise. In addition, unrelated functions are extracted using regular-expression-based function matching, which may fail to identify some functions with complex C/C++ syntax. These limitations mainly affect the Whole-File evaluation. The Function-Pair setting is less exposed to this threat because it relies on manually verified vulnerable functions and their corresponding patched versions.

\section{Conclusion}
This study presents a deployment-oriented framework for evaluating learning-based vulnerability detectors through pre-training feature representation analysis, within- and cross-dataset testing, and temporally separated evaluation on VentiVul. Our results show that the evaluated graph-based and CodeBERT-based features provide limited and inconsistent separable signals before classifier training, while trained DL models perform well mainly under matched benchmark conditions and degrade substantially across independently constructed datasets. On recent Linux and Chromium vulnerabilities, both trained DL models and off-the-shelf LLMs exhibit low detection effectiveness and limited ability to distinguish vulnerable functions from their patched counterparts. By combining Whole-File classification with Function-Pair evaluation metrics, this work demonstrates that benchmark performance and vulnerability coverage do not necessarily indicate transferable or patch-sensitive vulnerability detection.




\bibliographystyle{elsarticle-harv}
\bibliography{reference}

\end{document}